\pdfoutput=1
\documentclass[%
 aip,
 jmp,%
 amsmath,amssymb,
 reprint,%
]{revtex4-1}

\usepackage{graphicx}
\usepackage{dcolumn}
\usepackage{bm}
\usepackage{mathtools}
\usepackage{hyperref}
\usepackage{mathrsfs}
\usepackage{bbm}
\usepackage[toc,page]{appendix}

\hypersetup{
    colorlinks,
    citecolor=black,
    filecolor=black,
    linkcolor=black,
    urlcolor=black
}

\newcommand{\ii}{\mathbbm i}
\newcommand{\e}{\mathbbm e}
\newcommand{\1}{\mathbbm 1}
\newcommand{\dip}{\mathbbm P}

\begin{document}


\title[Gauge-Free Duality in Pure Square Spin Ice: Topological Currents and Monopoles]{Gauge-Free Duality in Pure Square Spin Ice: Topological Currents and Monopoles}

\author{Cristiano Nisoli}
 \email{\\ cristiano@lanl.gov \\ cristiano.nisoli.work@gmail.com}

\affiliation{ 
Theoretical Division, Los Alamos National Laboratory\\ Los Alamos, NM, 87545, USA
}%


\date{\today}

\begin{abstract}
We consider a pure square spin ice, that is a square ice where only nearest neighbors are coupled. A gauge-free duality between the perpendicular and collinear structure  leads to a natural description in terms of topological currents and charges as the relevant degrees of freedom. That, in turn, can be expressed via a continuous field theory where the discrete spins are subsumed into entropic interactions among charges and currents. This approach produces structure factors, correlations, and susceptibilities for spins, monopoles, and currents. It also generalizes to non-zero temperature the height formalism of the disordered ground state. The framework can be applied to a zoology of recent experimental results, especially realizations on quantum annealers and can be expanded to include longer range interactions.
\end{abstract} 

\maketitle

\tableofcontents\vspace{20mm}

\section{Introduction} 

Degenerate artificial square ice is perhaps the simplest two-dimensional system in which to study  disorder constrained by the ice rule\cite{lieb1967residual}, and its violations as monopole excitations\cite{ryzhkin2005magnetic,Castelnovo2008}.  
While spin ice  pyrochlores\cite{bramwell2020history,Ramirez1999,den2000dipolar,Bramwell2001} had opened new vistas in the study of geometric frustration and constrained disorder, more recently {\em artificial spin ices}---systems of interacting, magnetic nanoislands\cite{tanaka2006magnetic,Wang2006,Nisoli2013colloquium,heyderman2013artificial,skjaervo2019advances,ortiz2019colloquium}---have provided controllable platforms that can be characterized at the constituent level. 

Although the field has developed to include new forms of frustration and geometries, allowing for the realization of magnets often not found in nature\cite{Morrison2013,stamps2014artificial,nisoli2017deliberate,li2018superferromagnetism,saccone2019dipolar,makarova2021low,gliga2017emergent, stopfel2018magnetic}, and revealing novel phenomena absent in its crystal analogue\cite{skjaervo2019advances,schiffer2021artificial}, recent fabrication and characterization advances\cite{arnalds2012thermalized,farhan2013direct,kapaklis2014thermal} have brought back to the fore the austere simplicity of celebrated early models, such as kagome and square ices.


The square geometry of spin ice was among~\cite{tanaka2006magnetic} the first to be realized artificially~\cite{Wang2006}, when it was shown that non-ice rule~\cite{bernal1933theory} vertices are suppressed  after AC demagnetization~\cite{Wang2006,Nisoli2007,Nisoli2010}. 
However, because in nanoislands moments impinging  perpendicularly in the vertex interact more strongly than moments impinging collinearly, the degeneracy of the ice manifold is lifted and antiferromagnetic (AFM) vertices are  favored, leading to a phase transition in the Ising class\cite{Wu1969,anghinolfi2015thermodynamic} toward an ordered antiferromagnetic ground state\cite{Morgan2010,Porro2013,zhang2013crystallites,sendetskyi2019continuous}.

Soon after, M\"oller and Moessner~\cite{Moller2006}  proposed to offset the height of half of the nanoislands to regain the ice rule degeneracy~\cite{giauque1933molecular,Pauling1935}. The idea was recently realized\cite{perrin2016extensive,farhan2019emergent}. Meanwhile, ice rule degeneracy in square ice has been also demonstrated in rectangular lattices~\cite{Nascimento2012,ribeiro2017realization}, or via cleverly placed interaction modifiers placed in the vertices\cite{ostman2018interaction}, or nano-holes in a connected spin ice of nanowires~\cite{schanilec2019artificial}, or by rotation of the moments~\cite{macedo2018apparent}. Recently, a pure square ice, {\em i.e.}\ a square ice with only nearest neighbor interaction, was realized in a quantum annealer~\cite{king2021qubit}, and used to demonstrate purely entropic monopole interactions. Then, various different regimes can be achieved by tweaking specs and couplings, leading to antiferromagnetic states, line states, or ice manifolds, which lead to different spectral characterizations~\cite{rougemaille2021magnetic}.  

The scope of this work is to provide an unifying and expandable framework that can cover many different square ice systems around the ice rule degeneracy point~\cite{Moller2006,king2021qubit}, by considering {\em topological charges} and {\em currents} as the relevant degrees of freedom and subsuming the spin structure into effective, entropic interactions among them. This approach flows naturally from a gauge-free duality of the system which is absent in three dimensions (3D). 

We limit ourselves to {\em pure} square ice, that is a  sixteen vertex model where interactions are limited to spins within the same vertex~\cite{lieb1972inphase,Baxter1982,Wu1969}. We consider no long-range interactions, and thus no 3D-Coulomb (i.e.\ $1/r$) interaction among monopoles or currents (we have considered such case elsewhere~\cite{nisoli2020equilibrium}). However, because of the emergent nature of these objects, we show that they interact via 2D-Coulomb (i.e. $\sim \ln r$) entropic interactions.
 
 Within this model, we compute free energies, entropic interactions, structure factors, susceptibilities, correlations,  screening, relaxation dynamics. We discuss strengths and limits of this approach. We show new results, but also re-derive in a coherent framework results that were previously appreciated in similar systems through a variety of methods. These had included phenomenological approaches via coarse grained field,  height models, or  analogies with chemical physics approaches~\cite{henley2010coulomb,isakov2004dipolar,garanin1999classical,henley2005power,youngblood1981polarization,huse2003coulomb,henley2011classical,henley1997relaxation,bramwell2012generalized,twengstrom2020screening,jaccard1964thermodynamics,ryzhkin2005magnetic}. We also particularized some results that what we had already found on generic graphs~\cite{nisoli2020concept}.

\section{Square Ice and Its Gauge-Free Duality}
\label{dual}

There is in 2D  a  {\em gauge-free} duality absent in 3D pyrochlore ice. It is related to the rather gravid mathematical fact that in 2D a Helmholtz decomposition has no gauge freedom. That in turns follows from the fact that orthogonal directions are uniquely defined in 2D. It is amply used in 2D continuum theories, from fluid dynamics to  the $XY$ model~\cite{kosterlitz1973ordering}, and  is behind the entire edifice of complex analysis.

\subsection{Gauge-Free duality in 2D}

In 2D we can always write a  continuum vector field $\vec S$  in terms of longitudinal and transverse potentials  $h_{||}$, $h_{\!\perp}$, 
\begin{align}
\vec S = \vec S_{||}+\vec S_{\!\perp}=\vec \nabla h_{||} - \hat e_3 \wedge \vec \nabla h_{\!\perp},
\label{Helm}
\end{align}
 where $\hat e_3=\hat e_1 \wedge \hat e_2$, $\hat e_1, \hat e_2$ is an orthonormal basis of the plane, and we call $\vec S_{||},\vec S_{\!\perp}$ the longitudinal and perpendicular components of the field. Unlike in the 3D case, where the perpendicular part of $\vec S$ is the curl of a vector potential, there is no gauge freedom in Eq~(\ref{Helm}).  
 
 If we  define the {\em charge} and {\em current} distributions  of the field as
 \begin{align}
 q[\vec S] \vcentcolon &= -  \vec \nabla \cdot \vec S \nonumber \\
 i[\vec S] \vcentcolon &=\hat e_3 \cdot \vec \nabla \wedge \vec S,
 \label{tito}
 \end{align}
 then   
\begin{align}
q &= -\Delta h_{||} \nonumber \\
 i &=-\Delta h_{\! \perp}.
\label{rhoj}
\end{align}
If for a vector $\vec w$  we call  
\begin{equation}
^{\perp}\! \vec w \vcentcolon =\hat e_3 \wedge \vec w
\end{equation}
 the {\em perpendicular} of $\vec w$, we have then 
\begin{align}
q[\vec S] &= i[^{\perp} \!\vec S] \nonumber \\
 i[\vec S] &= - q[^{\perp} \! \vec S],
\label{duality2}
\end{align}
which expresses the duality between charges and currents, or longitudinal and perpendicular components of the field, under a $\pi/2$ rotation.

This duality has a discretized analogue in square spin ice.

\subsection{Square Spin Ice} 
 
Square spin ice (Fig.~1) is a set of classical, binary spins $\vec S_e$ aligned on the $N_e$ edges $e$ of a square lattice of $N_v=N_e/2$ vertices labeled by $v$. Spins form four vertex topologies often classified~\cite{Wang2006} as t-I, \dots, t-IV, where  t-I and t-II obey the ice rule~\cite{bernal1933theory, Pauling1935,Baxter1982,Moller2006} (i.e.\ have two spin pointing in, two pointing out). 
 
 In a degenerate square ice, ice rule vertices  are degenerate and energetically favored, and lead to a  ground state~\cite{lieb1967residual} of constrained disorder, called the {\em Ice Manifold}. The latter is {\em a Coulomb phase}~\cite{henley2010coulomb,isakov2004dipolar,garanin1999classical,henley2005power}, i.e\ a topological state labeled,  in lieu of an order parameter, by a {\em height field}~\cite{henley2011classical,lamberty2013classical}.  

\begin{figure}[t!]
\center
\includegraphics[width=.99\columnwidth]{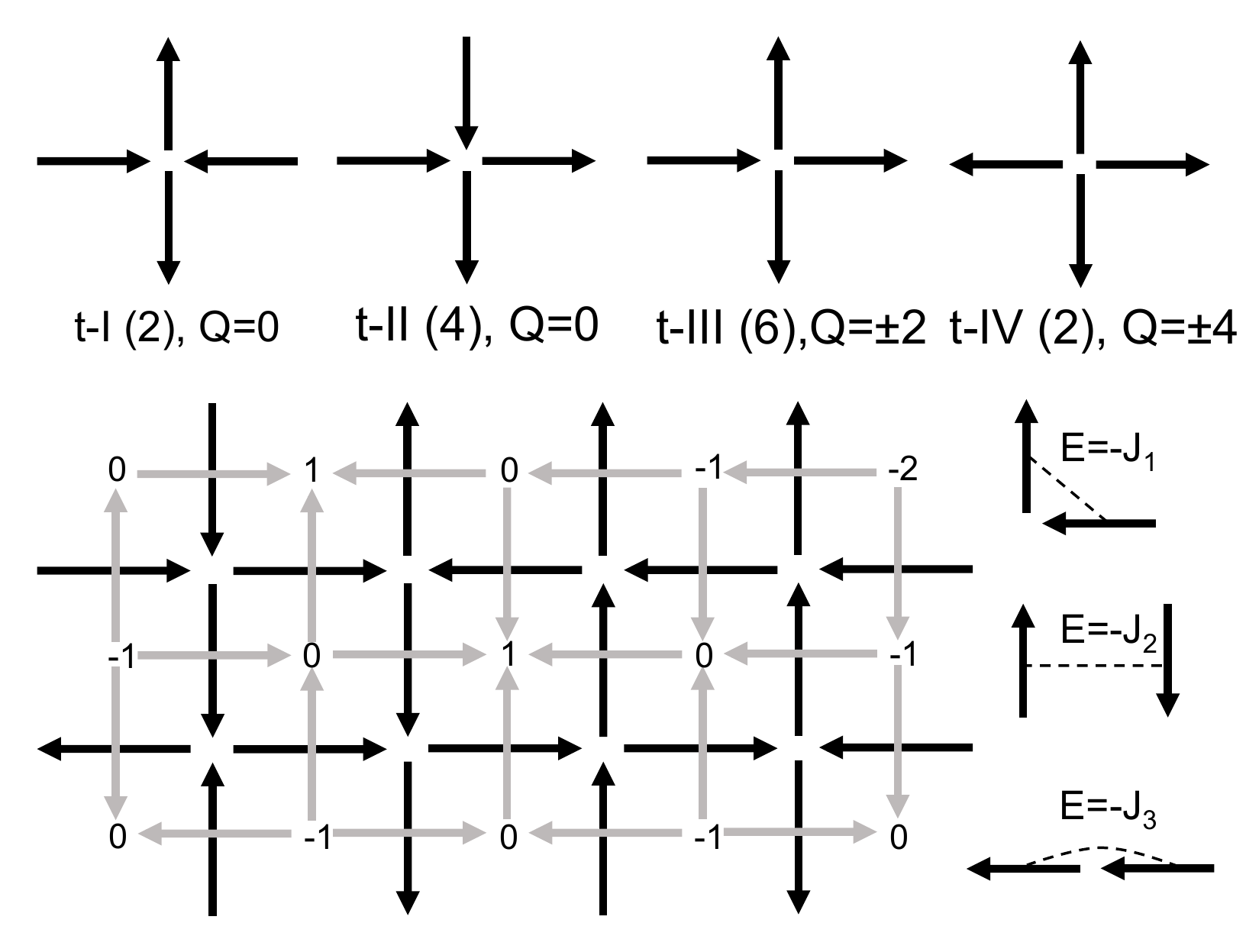}
\caption{Top: the sixteen vertices of square ice can be divided into four topologies, listed with degeneracy in parenthesis and topological charge. Below, an ice rule obeying configuration of $\vec S$ (black) and its height function $h_{\! \perp}$  built from $^{\perp}\! \vec{S}$ (gray). Also, the coupling constants among spins.}
\label{fig1}
\end{figure}

Consider $Q_v[S]$, the topological charge of the vertex $v$, defined as the number of spins pointing in the vertex minus those pointing out. Then, an  ice rule vertex $v$ has  $Q_v=0$. 

We can similarly define  the topological current $I_p[S]$ of a minimal square plaquette $p$, as the number of  spins pointing clockwise around the edge of the plaquette minus those pointing counterclockwise~\footnote{We call it current because a magnetization $\vec M$ generates an electrical current density $\vec j = \vec \nabla \wedge \vec M$}. 

For a spin configuration $\vec S$,  consider its perpendicular  configuration $^{\perp \!} \vec S$ (Fig.~1), for which old plaquettes are now  vertices and old vertices are now plaquettes. We have then
\begin{align}
Q_v[\vec S]&=I_v[^{\perp}\! \vec S], \nonumber\\
 I_p[\vec S]&= - Q_p[^{\perp} \! \vec S]
\label{duality}
\end{align}
as in Eqs.~(\ref{duality2}). 

For each  configuration of spins that obeys the ice rule configuration,   a unique (up to a constant) {\it height function} $h_{\!\perp }$ can be defined on the plaquettes  such that 
\begin{equation}
{^{\perp}\!\vec S_e }\cdot \hat{pp'}=h_{\! \perp p'}- h_{\! \perp p},
\label{heighperp}
\end{equation}
is true ($\hat{pp'}$ is the unit vector pointing from plaquette $p$ to $p'$ separated by the edge $e$ as in Fig.~1). This follows from the fact that    $^{\perp \!}S$ is ``irrotational'': the line-sum of spins $^{\perp \!} \vec S$ (grey in Fig~1) along a closed loop is zero.

 Similarly, if a spin configuration  has  zero topological current on each plaquette ($I_p[S]=0$ $\forall p$), a  height function $h_{|| v}$ can be defined on the vertices  by 
\begin{equation}
\vec S_e \cdot  \hat{vv'} = h_{|| v'}
-h_{|| v}
\label{heighpar}
\end{equation}
where $e$ is the edge connecting the vertices $vv'$. 

Thus, in the ice manifold there are no charges, currents are disordered, and  $^{\perp}\!S$ is the discrete gradient of $h_{\! \perp}$, which ``labels'' the disorder of currents in absence of charges. Conversely, in the current-free manifold,  $S$ is the discrete of $h_{||}$, which labels the disorder of charges in absence of currents. 

The Eqs.~(\ref{heighperp}, \ref{heighpar}) are thus the discrete analogous to  Eqs.~(\ref{Helm}) in the continuum. A significant difference  is that  they  are  well defined only in the charge free or current free state.   We will show in the section~\ref{correlations} how to generalize them to any spin ensemble, inclusive of monopole and current excitations, and thus at non-zero temperature.  

\subsection{Heuristic Entropy and Ice-Like Correlations}

The height formalism can usefully, if heuristically, describe  the pure ice manifold.  Height models are said to be in a ``rough'' (degenerate) of ``flat'' (ordered) phase, a jargon derived from the theory of the {\em roughening transition} historically associated with these models by various exact mappings~\cite{van1977exactly,chui1976phase}. 

The ice manifold of the  square ice, i.e.\ the six vertex model, is known to be equivalent to a dimer cover model~\cite{zinn2009six,Baxter1982} and thus in a rough phase~\cite{henley2010coulomb,henley2011classical,henley1997relaxation}.
A widespread\cite{van1977exactly,chui1976phase,henley1997relaxation,henley2011classical} though by no means rigorously justified ansatz ascribes to a configuration in the ice manifold an  entropy~\cite{raghavan1997new} that is quadratic in the height function. In our language we can write
\begin{equation}
{\cal S}[h_{\!\perp}]= -\frac{1}{2 \chi_0}\int \left(\vec \nabla h_{\! \perp} \right)^2 d^2x,
\label{hen}
\end{equation}
 where $ h_{\! \perp}$ is homogenized into a continuum field and $\chi_0$ is a positive uniform susceptibility (see later). 
(Clearly, the same is true for the current-free manifold by replacing $h_{\! \perp}$ with $h_{\! ||}$.) 

Eq.~(\ref{hen}) can be understood in terms of the zero temperature partition function
\begin{equation}
Z[\vec H]=\int \left[dh_{\! \perp}\right] \exp \left({{\cal S}[h_{\!\perp}]+ \hat e_3 \cdot\int \vec H \wedge \nabla h_{\! \perp}d^2x }\right)
\label{hen2}
\end{equation}
where $\vec H$ is an external field. 
 
Note that Eqs.~(\ref{hen},\ref{hen2}) are not obviously unproblematic in a 2D (and thus gauge-free) theory. In 3D we would be safe, as gauge invariance of the transverse part of the field forbids the proliferation of relevant operators at the fixed point (which, incidentally, is why an Higgs boson is needed in the standard model). Eq.~(\ref{hen}) merely happens to work in reproducing correlations that can also in part be computed exactly\cite{sutherland1968correlation,Baxter1982} (see also ref\cite{henley1997relaxation} and references therein for a discussion). 

A series of interesting deductions come from Eqs.~(\ref{hen},\ref{hen2}). For   heigh function correlations  one immediately finds in reciprocal space
\begin{equation}
\langle | \tilde h_{\! \perp}(k)|^2\rangle= \frac{\chi_0}{k^2}.
\label{heighcorr}
\end{equation}
From  that and the continuum limit of  Eq~(\ref{heighperp}) (or $\vec S = - \hat e_3 \wedge \vec \nabla h_{\!\perp}$) one obtains the spin correlator as~\footnote{We use dyadics: if  $\vec v$ is a  vector, the components of $\vec v \vec v$ are $[\vec v \vec v]_{ij} \vcentcolon= v_iv_j$; $\1$ is the unitary matrix.}
\begin{equation}
\chi_0^{-1}\langle \vec{\tilde{S}}^*(k) \vec{\tilde{ S}}(k)\rangle=  \frac{^{\perp}\! \vec k ^{\perp}\! \vec k}{k^2}=\1 -\frac{ \vec k \vec k}{k^2}
\label{spincorrheigh}
\end{equation}
  which is purely transversal. 
  
  Note that from Eq.~(\ref{heighcorr}),  in real space, correlations of the height function are { logarithmic}, or
\begin{equation}
\langle  h_{\! \perp}(x) h_{\! \perp}(y)\rangle= -\frac{\chi_0}{2\pi}  \ln(|x-y|).
\label{}
\end{equation}
Spin correlations in real space can be obtained from Eq.~(\ref{spincorrheigh}) or more easily as partial transversal derivatives (or $^{\perp}\vec \nabla =\hat e_3 \wedge \vec \nabla $) of Eq.~(\ref{heighcorr}), obtaining 
\begin{equation}
\chi_0^{-1}\langle  \vec S (x) \vec S(0)\rangle=\delta (x) \1  + \dip(x),
\label{heighcorr2}
\end{equation}
where $\dip(x)$ is the kernel of the dipole-dipole interaction in 2D, or
\begin{equation}
 \dip(x) =\frac{1}{2\pi} \left(\frac{\1}{x^2}-2\frac{\vec x \vec x}{x^4}\right).
\label{P}
\end{equation}
Analogously, in the case of pyrochlore spin ice the spins correlations are the kernel of the 3D dipolar interaction\cite{henley2010coulomb}.

The spin correlations are therefore algebraic, making the ice-manifold a critical phase of infinite correlation length. On the other hand the correlation length for currents is zero: from Eqs.~(\ref{heighcorr}, \ref{rhoj}) we have
\begin{equation}
\langle |\tilde i(k)|^2\rangle= {\chi_0}{k^2},
\label{ii}
\end{equation}
 which implies the infinitely localized screening of any pinned current. 

In terms of currents, from Eq.~(\ref{hen}) the entropy for the ice manifold can be rewritten as 
\begin{equation}
{\cal S}[i]= \frac{1}{2\chi_0}\int d^2 x~\! i(x)\ln(x-y) i(y)~\!d^2y,
\label{Slog}
\end{equation}
i.e. as a pairwise 2D-Coulomb interaction among the currents. 

Our phenomenological picture is thus the following: in the ice manifold charges are absent, disorder can be labeled by currents and their 2D-Coulomb mutual interaction determines the entropy.

From Eqs.~(\ref{hen}, \ref{Helm}), the entropy of a configuration in the ice manifold,  can be written in terms of $\vec S$ as
\begin{equation}
{\cal S}[\vec{S}]=- \frac{1}{2\chi_0} \int \vec{S}^2 d^2x,
\label{hen3}
\end{equation}
with the constraint \hbox{$\vec \nabla \cdot \vec S=0$}. 

Equation~(\ref{hen3}), unlike our previous formulas, is also valid in 3D. There, it has been appreciated as {\em Jaccard entropy} in water ice~\cite{jaccard1964thermodynamics,nagle1966lattice,nagle1978configurational,nagle1979theory,ryzhkin1997configurational}, and later in pyrochlore spin ice~\cite{huse2003coulomb,isakov2004dipolar,henley2005power,henley2010coulomb} as necessary to produce purely transverse correlations in the ice manifold. We see therefore  that in square ice the Jaccard entropy describes in fact a 2D-Coulomb interaction among disordered currents. The uniform susceptibility $\chi_0$  is thus related to the so-called  $\Phi$ constant~\cite{ryzhkin1997configurational}. 

We note that in the previous deductions we have taken some cavalier liberties with the boundary terms. With collaborators, we have already shown how fixing the boundaries can induce a net charge in the bulk via a geometrical expression of the Gauss's Law~\cite{king2021qubit}. Taking the orthogonal of the spin distribution, that implies a net current in the bulk if a current is present on the boundaries. In a future work we will consider these interesting topological effects for currents at the boundaries.



In the following we will show how to deduce a field theory for charges and currents in which the heuristic Eqs.~(\ref{Helm}, \ref{rhoj}) make sense, the intuitive height function formalism [Eqs.~(\ref{hen}-\ref{hen3})] finds a solid ground, and it is generalized for $T>0$. 


\section{Energy and States}

The following  Hamiltonian 
\begin{align}
{\cal H}[Q,I]=\frac{\epsilon}{2}\sum_v Q_v^2 +\frac{\kappa}{2}\sum_p I_p^2, 
%
\label{H-n}
\end{align}
 reflects the current-charge duality by placing a cost or advantage on topological currents and monopoles ($\epsilon$ and $\kappa$ are energies). 
In terms of an Ising model, it is equivalent to a $J_1, J_2, J_3$ model where  (Fig.~1): $J_1=\epsilon-\kappa$ , $J_2=-\kappa$, $J_3=\epsilon$. By the duality, the symmetry by orthogonalization corresponds to $\epsilon \leftrightarrow \kappa$.
The Hamiltonian  describes various cases, often close to the experimental reality. 

If $\kappa=0$ and $\epsilon>0$,  the ground state is the ice manifold of Fig.~1 (black arrows). Equivalently, by gauge-free duality, if $\epsilon=0$ and $\kappa>0$, the ground state is an extensively degenerate ice manifold for $^{\perp}\!\vec S$, or the grey arrows in Fig.~1.

 If $\kappa=0$ and $\epsilon<0$, the ground state is the charge full state, i.e. the ordered, antiferromagnetic tessellation of t-IV vertices.  If $\epsilon=0$ and $\kappa<0$ the ground state is the current full state, i.e. the ordered, antiferromagnetic tessellation of t-I vertices, which is the orthogonal of the charge full state.

For $\kappa>0$, $\epsilon>0$,  both charges and currents are  suppressed. Because $J_1<J_3$, ferromagnetic t-II vertices are promoted over t-I. Because $J_2<0$, t-II vertices want to align and the ground state is the four-fold ferromagnetic state, made of t-II vertices ferromagnetically  aligned. (More loosely: as the ground state is current-free and charge-free, we have  $\Delta h_{||}=\Delta h_{\! \perp}=0$ which implies a uniform $\vec S$.)   
This case has not been investigated experimentally, though it can certainly be realized in a quantum annealer~\cite{king2021qubit}. It might approximate, however,  experimental situations where  t-II vertices can be  favored\cite{perrin2016extensive,ostman2018interaction,perrin2019quasidegenerate}, leading to a line state that is disordered but of sub-extensive entropy. 
 
For $\kappa<0, \epsilon>0$,  the ice rule is enforced at low $T$, currents are promoted, $J_1>J_3$, and the ground state is an ordered antiferromagnetic (AFM) tessellation of t-I vertices. Large $|\kappa|$  describes early square ice realizations\cite{Wang2006}. Small $|\kappa|$ might approximate spin ices that are designed so that ice rule vertices are degenerate~\cite{Moller2006,perrin2016extensive} but where the dipolar interaction still favors closed magnetic fluxes and thus  promotes topological currents and an ordered ground state.

When $\kappa=\epsilon$, we have $J_1=0$ and  the set of vertical and horizontal arrows become two decoupled systems. When $\kappa=\epsilon>0$ each subsystem is ferromagnetic and the ground state is the four-fold fully polarized state. When $\kappa=\epsilon<0$, each subsystem is antiferromagnetic and the ground state is a four-fold antiferromagnetic state, which includes the two orientations of the charge-full state and the two orientations of the current-full state (one is the orthogonal of the other, as expected on the symmetry line $\kappa=\epsilon$).

We will not study here the full monopole-currents model of Eq~(\ref{H-n}), which leads to a rich phase diagram in the  $\beta \epsilon \times \beta \kappa$ plane (where $\beta = 1/T$ and $T$ is temperature measured in units of energy) to be compared with other models~\cite{levis2013thermal,Wu1969}. 
We will  consider only $\epsilon>0$ and $|\kappa|/\epsilon$ small 
and  investigate how ice manifold features are retained by small perturbations around the {\em spin ice point}  $\kappa=0$. 

\section{Field Theory: Exact Results}

Because charges and currents represent an emergent description of pure square  ice, we deduce a field theory for which they are the relevant degrees of freedom.

We generalize  our previous approach for general graphs~\cite{nisoli2020concept,nisoli2020equilibrium} to include currents.
 The  partition function from Eq.~(\ref{H-n}) reads
\begin{align}
Z&=\sum_{S} \exp \left(-\beta {\cal H} \right) \times  \nonumber \\ 
& \exp \! \left(\beta \sum_e \vec S_e\cdot \vec H_e +\beta \sum_v V_{q,v} Q_v + \beta \sum_p V_{i,p} I_p\right),
\label{Z-n}
\end{align}
and it is the generator of correlations
\begin{align}
&\langle \vec S_{e_1} \dots \vec S_{e_n}\rangle=\partial_{\beta \vec H_{e_1} \dots \beta \vec H_{e_n}} \! \! \ln Z \nonumber \\  
&\langle Q_{v_1} \dots Q_{v_n}\rangle=\partial_{ \beta V_{q,v_1}} \dots \partial_{\beta V_{q,v_n}} \! \! \ln Z \nonumber \\ 
&\langle I_{p_1} \dots  I_{p_n} \rangle=\partial_{\beta V_{i,p_1}} \dots \partial_{\beta V_{i,p_n}} \! \! \ln Z.
\label{corr1}
 \end{align}
The fields $\vec H, V_q, V_i$ are measured in units of energy.

To obtain a continuum field theory we insert  in the sum of (\ref{Z-n}) the tautology 
\begin{align}
1&=(2\pi)^{-2N_v}\prod_v \int   dq_v d\phi_v\exp\left[\ii \phi_v\! \! \left(q_v-Q_v\right) \right] \nonumber \\
& \times \prod_p \int   di_p d\psi_p\exp\left[\ii \psi_p\left (i_p-I_p\right) \right]
\end{align}
and then sum over the spins, obtaining
\begin{equation}
Z=\int \left[dq di\right]  \tilde \Omega[q,i]  \mathbbm{e}^{-\beta{\cal H}[q,i] + \sum_v  q_v V_{q,v}+ \sum_p  i_p V_i,p}
\label{Z2-n}
\end{equation}
where $\left[dq di\right]=(2\pi)^{-N_v} \prod_v dq_v \prod_p d i_p$. 
$\tilde \Omega[q,i]$ is a generalized density of states for  $q_v, i_p$,  given by
\begin{align}
\tilde \Omega[q,i]=\int [d\phi d\psi] \Omega[\phi, \psi]\e^{ \sum_v  q_v \ii \phi_v + \sum_p  i_p \ii \psi_p },
\label{Omegatilde-n}
\end{align}
and is therefore the functional Fourier transform of 
\begin{equation}
\Omega[\phi, \psi]=2^{N_e}{\prod_{{\hat{vv'}}}}\cosh \left(-\ii \nabla_{vv'}\phi -\ii \nabla_{pp'}\psi+  \beta H_{vv' } \right).
\label{Omega-n}
\end{equation}
%
(The product runs on all the edges $e={\hat{vv'}}$ once, and  $\nabla_{vv'}\phi \vcentcolon=  \phi_{v'}- \phi_{v}$,  $\nabla_{pp'}\psi \vcentcolon= \psi_{p'}- \psi_{p}$, $ H_{vv' } \vcentcolon=  \vec H_e\cdot \hat{vv'}$, while $\hat{pp'}=-\hat e_z \wedge \hat{vv'}$.) 

Note that by construction $\langle Q_{v_1} \dots Q_{v_n}\rangle=\langle q_{v_1}\dots  q_{v_n}\rangle$, $\langle I_{p_1} \dots I_{p_n}\rangle=\langle i_{p_1}\dots  i_{p_n}\rangle$. Note also that the use of $\vec H$, $V_q, V_i$ is superabundant: $V_q \to V_q + V'_q, V_i \to V_i + V'_i$ is equivalent to $\vec H \to \vec H+\vec \nabla V'_q - \hat e_3\wedge  \nabla V'_i$, but we keep it because it is useful.

We have  gone from binary variables to a theory of continuous emergent topological charges and currents constrained by an entropy 
\begin{equation}
S[q,i]=-T\ln \tilde \Omega[q,i]
\end{equation}
 which conveys the effect of the underlying spin ensemble. Equivalently, in the language of field theory,  charges  $q_v$ and currents $i_p$ interact  {\em entropically} via the fields
\begin{align}
&V_q^e= \ii T  \phi   \nonumber \\
&V_i^e= \ii T  \psi,
\label{Ve}
\end{align}
of  generalized free energy  
 \begin{equation}
 {\cal F}[\phi, \psi]=-T\ln \Omega[\phi, \psi].
 \end{equation} 

Note that $ {\cal F}[\phi, \psi]$ can have imaginary values. As a consequence, though the variables $\phi, \psi$  are themselves real,   their expectation values $\langle  \phi \rangle$, $\langle  \psi \rangle$ are  imaginary, and thus the entropic fields $\langle V_q^e\rangle, \langle V_i^e\rangle$ are real. Indeed, by integrating over $q$ in Eq~(\ref{Z2-n})  and applying the second and third equation in (\ref{corr1}), one finds
\begin{align}
&\langle V_{q}^e\rangle +V_{q} =\epsilon \langle q \rangle, \nonumber \\
&\langle V_{i}^e\rangle+V_{i} =\kappa \langle i \rangle,
\label{eos0}
\end{align}
 when $\kappa>0$. 
 Similar Gaussian gymnastics also prove that
  \begin{equation}
 \langle S_{vv'}\rangle=\langle \tanh \left(\beta H_{vv'}-\ii \nabla_{vv'}\phi - \ii \nabla_{pp'}\psi  \right) \rangle.
 \label{M}
 \end{equation}
 We thus see that while $\ii\phi, \ii\psi$ correlate charges and currents, the vector fields $- \ii T \nabla_{vv'}\phi$, $-\ii T \nabla_{pp'}\psi$ correlate spins that would otherwise be trivially paramagnetic. 
 
 In the spirit of field theory, we could therefore say the following:  $V_q^e$ is an entropic potential acting on charges and $V^e_i$ on currents, and correspondingly there is an entropic field $\vec B^e$ acting on the spins which  in the long wavelength approximation is given by
\begin{align}
 \vec B^e = \vec B^e_{||}+\vec B^e_{\!\perp}=- \vec \nabla V_q^e - \hat e_3 \wedge \vec \nabla V_i^e.
\label{Be}
\end{align}
%

\section{Field Theory: Approximations}
 
\subsection{High Temperature}

 When $\epsilon >0$ and  $\kappa >0$ (see Section \ref{kneg} for $\kappa\le 0$) integrating over charges and currents in Eq.~(\ref{Z2-n})  returns Gaussians in the fields, of variance \, $\langle \phi^2 \rangle =\epsilon/T$, $\langle \psi^2 \rangle =\kappa/T$: as temperature increases, the entropic fields mediating spin correlations become smaller, as one would expect. In fact,  
in  the infinite temperature limit (but with $\beta H$ constant), the Gaussian distributions in the field become Dirac deltas, and  Eq.~(\ref{Z2-n}) becomes the standard ``paramagnetic'' partition function 
\begin{equation}
 Z=2^{N_s}\prod_{ \hat{vv'} } \cosh\! \! \left( \beta H_{\hat{vv'} }\right).
\end{equation}
 Note that for  $H=0$, the equation above returns the correct entropy per spin at infinite temperature, $s=\ln2$. 

In the high $T$ limit, we can linearize Eq.~(\ref{M}) and write, in the long wavelength limit,
\begin{align}
\langle \vec S \rangle =  \beta \left( \vec H + \langle \vec B^e \rangle\right).
\label{Helm2}
\end{align}
Then, from $q= - \vec \nabla \cdot \vec S$, $j=\hat e_3 \cdot \vec \nabla \wedge \vec S$, and Eqs.~(\ref{eos0}) we find screened Poisson equations for $q$, $i$ 
\begin{align}
-\xi_{||}^2 \Delta q +q&= \beta\nu q_{\text{ext}}, \nonumber \\
 -\xi_{\perp}^2 \Delta i +i&= \beta\nu i_{\text{ext}}
 \label{screen}
\end{align}
(here ${\nu} q_{\text{ext}}=\vec \nabla \cdot \vec H$, where $q_{\text{ext}}$ is the external charge, $ \nu i_{\text{ext}}=\vec \nabla \wedge \vec H$, where $i_{\text{ext}}$ is the external current, $\nu$ is an energy) and therefore  ``correlation lengths'' \footnote{When monopoles interact via a $1/r$ law,  $\xi_{||}$ is no longer a correlation length, as the monopole-monopole interaction destroys the screening at least in principle, as we have shown elsewhere~\cite{nisoli2020equilibrium}.} as
\begin{align}
 \xi^2_{||} &={\epsilon/T} \nonumber \\
 \xi^2_{\! \perp} &=\kappa/T.
\label{corrlen}
\end{align}
This expression for $\xi_{||}$ was already appreciated in other works, for different geometries, via different means\cite{garanin1999classical,henley2005power} and we had generalized it on a graph~\cite{nisoli2020concept}. 
Note that for $\kappa<0$, $ \xi_{\! \perp}$ would be imaginary, pointing to a periodicity typical of the AFM ensemble, as we will discuss in section \ref{kneg}. Instead when $\kappa=0$ the integral over $i$ returns a functional delta function on $\psi$, and $\psi$ disappears from the equations, which for all purpose is equivalent to taking $\xi_{\perp}=0$.  

\subsubsection{Effective Energies and Entropic Interactions}

 We now  explore these heuristic deductions more precisely. Because we are interested  in the monopole liquid below the ice manifold threshold $T\simeq 2 \epsilon$ but above other possible low $T$ transitions, 
 a high $T$ approximation is a good starting point. Since it corresponds to small entropic fields, we  expand $\ln \Omega[\phi, \psi]$ at quadratic order and Fourier transform via 
%
$g_x=\int_{BZ}  \tilde g(\vec k)\e^{-\ii \vec k \cdot x }{d^2k}/{(2\pi)^2},$
%
 where BZ is the Brillouin Zone, 
 $g$ is a generic field, and $x=v,p,l$  represents vertices, edges or plaquettes. 
 
 We obtain the approximated partition function at zero loop
\begin{align}
Z_2=\int [dq di][ d\phi d\psi]\exp\! \left(-\int_{\text{BZ}} \beta {\cal H}_2(\vec k) \frac{d^2k}{(2\pi)^2} \right) \nonumber \\
\label{Z2-n2}
\end{align}
where 
\begin{align}
{\cal H}_2[\tilde q, \tilde i, \tilde \phi, \tilde \psi]&  = \frac{\epsilon}{2} \left|\tilde{q}\right|^2 + \frac{\kappa}{2} |\tilde{i}|^2 + \frac{T \chi_0}{2 }  \gamma^2 \! \! \left(| \tilde   \phi |^2 + | \tilde \psi |^2 \right)  \nonumber \\
& - \ii T \! \left(\tilde{q}^*\tilde{\phi} + \tilde{i}^*\tilde{\psi}\right) \nonumber \\
& + \ii  \chi_0 \!  \left(^{\perp}\!\vec{\gamma} \tilde{\psi}^*  -   \vec \gamma \tilde{\phi}^* \right)\cdot \vec{\tilde{H}}   - \chi_0 \frac{\beta}{2}  \left| \vec{\tilde{H}}\right|^2
\label{f2-n}
\end{align}
is an effective Hamiltonian for the new variables and the vector $\vec \gamma$ has components 
\begin{equation}
\gamma_{\alpha} \vcentcolon = 2\sin(k_{\alpha}/2)
\end{equation}
for $\alpha=x,y$, and contains informations on the square symmetry of the lattice [in the long wavelength limit: $\vec \gamma \simeq \vec k+ O(k^3)$]. 

The first line of Eq.~(\ref{f2-n}) contains the free energies for the uncoupled charge, currents and entropic fields.  

The second line contains the  coupling between charges, currents and their entropic fields:
importantly, no $\phi  \psi$ cross term survives at quadratic order in the high $T$ approximation and charge and currents are independent at second order, leading to so-called magnetic fragmentation \cite{brooks2014magnetic,petit2016observation,canals2016fragmentation}, that is the decoupling between charge-full and current-full ensembles.

The third line contains the coupling with the external field $\vec H$ (we neglect here $V$). Because  $-i\vec\gamma \cdot \vec{\tilde w}$, $-i\vec\gamma \wedge \vec{ \tilde w}$ are the generalized divergence and curl respectively on the lattice in  moment space for a generic field $\vec w$,  the entropic fields for currents (resp. charges) couple to the curl (resp. divergence) of the external field. 

The quantity $\chi_0$, already encountered in the previous section on the opposite limit ($T=0$) is the uniform susceptibility. In Eq.~(\ref{f2-n}) it is $\chi_0=1$, but we include it nonetheless in the equation because it can be $\chi_0 \ne 1$ at low temperatures (see below).

%

Integrating $Z_2$ over $\tilde \phi, \tilde \psi$ when $\tilde H=0$ returns the effective free energies for $q$ and $i$ in absence of external field, which are decoupled at second order:
\begin{align}
&{\cal H}_2 [q, i] = {\cal H}_2 [q] +{\cal H}_2 [i],
\label{f2rho}
\end{align}
where the two terms can be written as
\begin{align}
\beta{\cal H}_2 [q]& = \frac{1}{2 \chi_0}\! \! \left({\xi^2_{||} }  +\gamma^{-2} \right) |\tilde{q}|^2 \nonumber \\
\beta {\cal H}_2 [i]& = \frac{1}{2 \chi_0}\! \! \left({\xi^2_{\perp} } +\gamma^{-2} \right) |{\tilde i}|^2.
\label{f2rho2}
\end{align}
%

In field theory language, the first terms in Equations~(\ref{f2rho2}) are the ``masses'' of the monopole or of the current. The second impliy that the underlying spin ensemble mediates a pairwise entropic interaction among charges (or currents) which  is the 2D-Coulomb potential. In real space  at large distances (where $\gamma^2\simeq k^2$) the entropic interactions for charges and currents are 
\begin{align}
V_q^e(v-v')\simeq -\frac{T}{\chi_0} \frac{q_v q_{v'}}{2\pi }\ln|v-v'| \nonumber \\
V_i^e(p-p')\simeq -\frac{T}{\chi_0} \frac{ i_p i_{p'}}{2\pi}\ln|p-p'| .
\label{Ve}
\end{align}

The origin of entropic interactions is in the emergent nature of charges and currents, which exist in a spin vacuum. An assignation of charges and/or currents to the system changes the number of ways in which the underlying spin ensemble can be compatibly arranged. Thus the entropy associated to a fixed distribution of charges and currents depends on their mutual position and distance. While this is obvious, it is not obvious that changes in entropy can be written via pairwise terms, as shown above, leading to an effective pairwise interaction. We see that subsuming the effect of the underlying spins leads to a ``2D electrodynamics" formalism for charges and currents. 

Integrating instead over $\tilde q, \tilde i$ in Eq.~(\ref{})

%
%
%
%

\subsubsection{Correlations, Susceptibilities, and Height Functions}
\label{correlations}
\begin{figure}[t!]
\center
\includegraphics[width=.99\columnwidth]{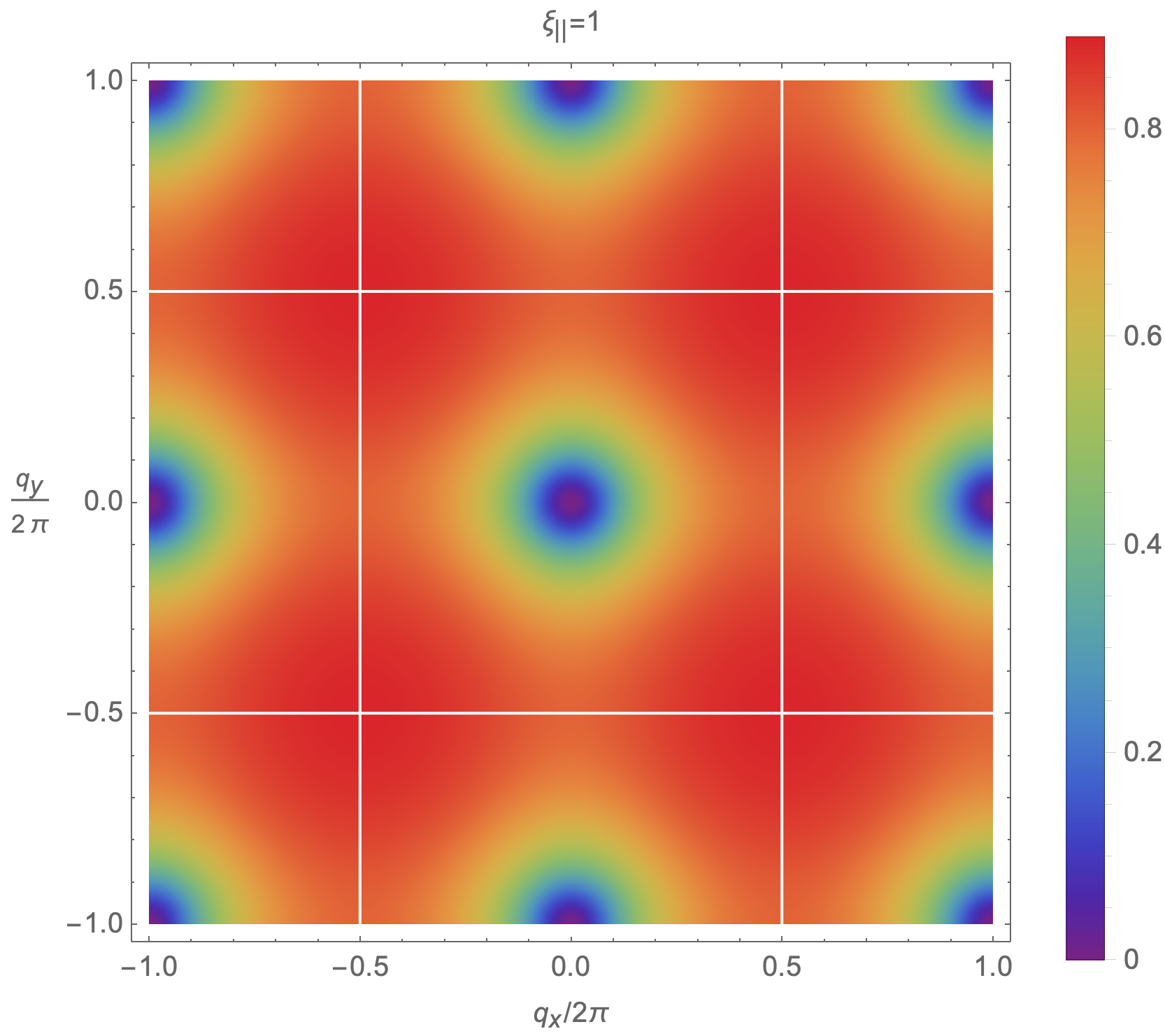}\vspace{5mm}
\includegraphics[width=.99\columnwidth]{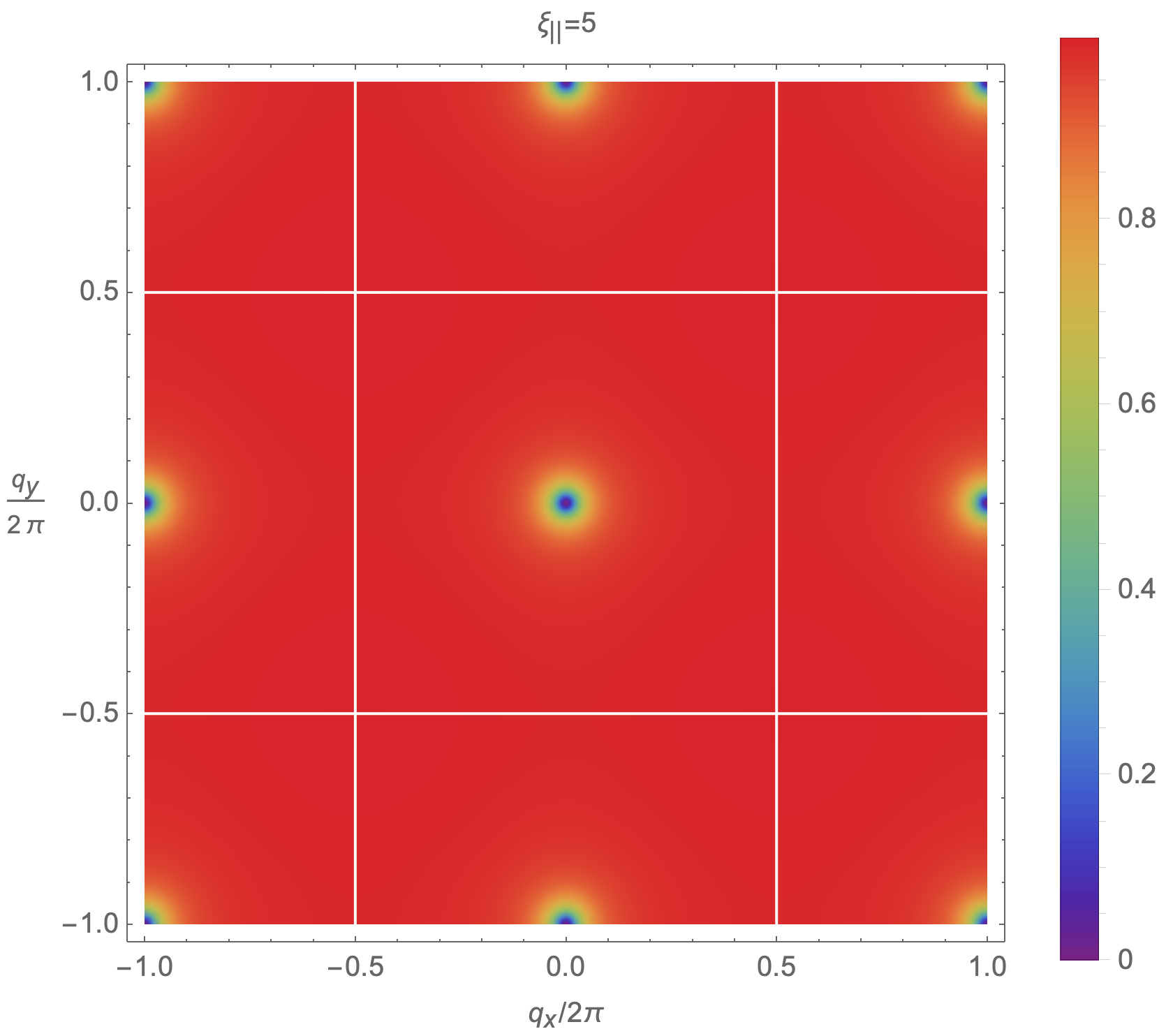}

\caption{Plots of the charge form factor $\langle | \tilde  q(\vec {k}) |^2\rangle \xi_{||}^2/\chi_0$ for $\xi_{||}=1$ (top) and $\xi_{||}=5$ (bottom). White lines denote the Brillouin zone. Identical plots hold for $\langle | \tilde  i(\vec {k}) |^2\rangle \xi_{\! \perp}^2/\chi_0$}.
\label{rho2fig}
\end{figure}

From Eq.~(\ref{f2rho}) and equipartition we obtain the correlations for charges and currents 
\begin{align}
&\langle | \tilde  q(\vec {k}) |^2\rangle = \gamma(\vec k)^2 \tilde \chi_{||}(k) \nonumber \\
&\langle | \tilde i (\vec {k}) |^2\rangle = \gamma(\vec k)^2 \tilde \chi_{\!\perp} (k),
\label{rhocorr}
\end{align}
where $\tilde \chi_{||}(k), \tilde \chi_{\! \perp} (k)$  are defined as
\begin{align}
 \tilde \chi_{||,\perp}(k)=\frac{\chi_0}{1 + \xi_{||,\perp}^2 \gamma(\vec k)^2}, \nonumber \\
\label{taus}
\end{align}
Correctly, from Eq.~(\ref{rhocorr}) we  have$\langle | \tilde  q(\vec {0}) |^2\rangle=0$, since $\langle | \tilde  q(\vec {0}) |^2\rangle= \langle \left( \sum_v  q_v \right) ^2\rangle=0$ (and same for currents). Figure~\ref{rho2fig} shows a heat map for $\langle | \tilde  q(\vec {k}) |^2\rangle$.

Furthermore, if $\langle q^2\rangle$ is the average charge per vertex, or $\langle q^2\rangle= \frac{1}{N_v}\sum_v \langle   q_v^2  \rangle$, then 
\begin{equation}
\langle q^2\rangle =\int_{\text{BZ}} \langle | \tilde  q(\vec {k}) |^2\rangle \frac{d^2k}{(2\pi)^2}.
\label{q2}
\end{equation}
 From it, and Eq.~(\ref{rhocorr}), we obtain $\langle q^2\rangle \to 4$  for $T \uparrow \infty$, which is correct. Indeed $\langle q^2\rangle = 4$  is the value deducible from a multiplicity argument ($2^2/2+4^2/8=4$). Because of the duality, the same is true for currents.

 Note that $\tilde \chi_{||}(k), \tilde \chi_{\! \perp} (k)$ are the the longitudinal and perpendicular static susceptibilities (multiplied by $T$, thus accounting already for the Curie-Weiss law) for  ${\tilde S}_{||}=\vec{\tilde S} \cdot \hat \gamma$ (the longitudinal,  charge-full, current-fee part of the magnetization) and ${\tilde S}_{\! \perp}=\vec{\tilde S} \cdot ^{\perp \! \!}\hat \gamma$ (the perpendicular, charge-free, current-full magnetization), respectively~\footnote{$\hat \gamma$ is  the unit vector of $\vec \gamma$}. Indeed, by integrating  the Gaussian integral in Eq.~(\ref{Z2-n2}), and remembering that $ \hat \gamma_{\alpha} \hat \gamma_{\alpha'} + {^{\perp} \! \hat \gamma_{\alpha}} \! \!{^{\perp} \! \hat \gamma_{\alpha'}} = \delta_{\alpha\alpha'}$, one obtains the total free energy as a function of the external field at lowest order as
\begin{align}
\beta{\cal F}_2 [ H ] = -\frac{1}{2} (\beta \vec{\tilde H}^*)\cdot \left( {\hat \gamma \hat \gamma}{ \tilde \chi_{||}} +  { {^{\perp} \! \hat \gamma} {^{\perp} \! \hat \gamma}}{\tilde \chi_{\! \perp}}\right)\cdot (\beta \vec{\tilde H}).
\label{ftot}
\end{align}
 From it, the spin correlations are
\begin{equation}
\langle \tilde S^*_{\alpha}(\vec k) \tilde S_{\alpha'}(\vec k)\rangle=  \hat \gamma_{\alpha} \hat \gamma_{\alpha'} \tilde \chi_{||}+   {^{\perp} \! \hat \gamma_{\alpha}} \! \!{^{\perp} \! \hat \gamma_{\alpha'}}\tilde \chi_{\! \perp}.
\label{spincorr}
\end{equation}
In the limit \hbox{$T\to \infty$}, Eq.~(\ref{spincorr}) correctly returns   \hbox{$\langle \tilde S_{\alpha}(\vec k) \tilde S_{\alpha'}(\vec k)\rangle \to \delta_{\alpha \alpha'}$}, or uncorrelated spins. 

Equation~(\ref{spincorr}) also implies that the magnetic susceptibilities in reciprocal space can be obtained from experimentally measured spin spin correlations as
\begin{align}
\tilde \chi_{\! \perp}(\vec k) & = ^{\perp} \! \! \hat \gamma \cdot \langle \vec{\tilde{S^*}}(\vec k) \vec{\tilde{S}}(\vec k) \rangle_{\text{exp}} \cdot^{\perp} \! \! \hat \gamma \nonumber \\
\tilde \chi_{||}(\vec k) & = \hat \gamma \cdot \langle \vec{\tilde{S^*}}(\vec k) \vec{\tilde{S}}(\vec k) \rangle_{\text{exp}} \cdot \hat \gamma
\end{align}
and similarly, from  (\ref{rhocorr}) 
\begin{align}
&\langle | \tilde  q(\vec {k}) |^2\rangle = \vec{\gamma}\cdot \langle \vec{\tilde{S^*}}(\vec k) \vec{\tilde{S}}(\vec k) \rangle \cdot \vec{\gamma}, \nonumber \\
&\langle | \tilde i (\vec {k}) |^2\rangle = ^{\perp}\! \!\vec{\gamma}\cdot \langle \vec{\tilde{S^*}}(\vec k) \vec{\tilde{S}}(\vec k) \rangle \cdot ^{\perp}\!\!\vec{\gamma}.
\label{rhocorrS}
\end{align}

Finally, from the spin correlations we can obtain the magnetic structure factor
\begin{equation}
\Sigma_m(\vec k)=^{\perp}\! \!\hat{k}\cdot \langle \vec{\tilde{S^*}}(\vec k) \vec{\tilde{S}}(\vec k) \rangle \cdot ^{\perp}\!\!\hat{k},
\label{sigma}
\end{equation}
which at small $k$ corresponds to $\tilde \chi \!_{ \perp}$.

Taking the long wavelength approximation, Eq.~(\ref{spincorr}) implies  the following effective free energy for the coarse grained spins 
\begin{equation}
\beta {\cal F}_2[S]=\int_{\text{BZ}} \! \left(\frac{1}{2 \tilde \chi_{||}} {\tilde S}_{||}^*\cdot {\tilde S}_{||} + \frac{1}{2 \tilde \chi_{\!\perp}} {\tilde S}_{\!\perp}^* \cdot {\tilde S}_{\!\perp}\right) \frac{d^2k}{(2\pi)^2}.
\label{FS}
\end{equation}
which in  the long wavelength limit becomes
\begin{equation}
\beta {\cal F}_2[S]= \frac{1}{2 \chi_0}\int \! \left[ {\xi^2_{||}}\left(\vec \nabla \cdot \vec S\right)^2 + {\xi^2_{\! \perp}}\left(\vec \nabla \wedge \vec S\right)^2 +   S^2\right] d^2x.
\label{FSr}
\end{equation}
The first two terms in the previous equation are energetic. 
The third is the Jaccard entropy~\cite{jaccard1964thermodynamics,ryzhkin2005magnetic} mentioned above.  In the ice manifold it  is the only surviving term, since $\xi_{\!\perp}=0$ and $\nabla \cdot \vec S=0$, thus returning Eq.~(\ref{hen3}), previously found heuristically.

Equation (\ref{FSr}), when $\xi_{\!\perp}=0$ reduces to the functional found elegantly  via methods of chemical physics for pyrochlore ice by Bramwell~\cite{bramwell2012generalized,twengstrom2020screening}. Indeed, when expressed in terms of spins $\vec S$ rather than currents and charges $i,q$, the two formalisms are expected to coincide.

Expressing the coarse grained $\vec S$  via height functions as in Eq.~(\ref{Helm}) we obtain the free energy for the heigh functions in the long wavelength limit, at quadratic oder:
\begin{align}
 \beta{\cal F}_2[h_{||},h_{\!\perp}] &=\frac{1}{2}\int \left[ \left( \vec \nabla h_{||}\right)^2 +\xi_{||}^2 \left( \Delta h_{||}\right)^2 \right]d^2x \nonumber \\
 &+\frac{1}{2}\int \left[  \left( \vec \nabla h_{\! \perp}\right)^2  +\xi_{\perp}^2 \left( \Delta h_{\! \perp}\right)^2 \right]d^2x.
\label{Fh}
\end{align}
When  $\xi_{\perp}=0$,  Eq.~(\ref{Fh})   represents the  generalization for $T>0$ of the heuristic Eq.~(\ref{hen}). When $T=0$ $h_{||}$ is constant  and Eq.~(\ref{Fh}) reduces to  Eq.~(\ref{hen}). Moreover, in fact,  Eq.~(\ref{Fh}) {\it tends} to  Eq.~(\ref{hen}) when $T\to 0$. For $T>0$ monopoles appear, and thus  the longitudinal height function is also present. 

\subsection{Low Temperature}
\label{LT}

When  $T\downarrow 0$, fluctuations of the entropic should fields diverge. One could then perform a proper study of the higher orders expansion of $\Omega(\phi,\psi)$ in terms of Feynman diagrams. Because of the complex shape of $\Omega(\phi,\psi)$ such program is  challenging. It might also be uninteresting in real systems.

That is because at very small $T$  our model becomes insufficient for real systems. Firstly, in magnetic systems long range interactions become relevant at low temperature, and it can induce ordering at equilibrium~\cite{melko2001long}. Secondly, regardless of the range of interaction,  creation of monopoles is suppressed  when $T\ll \epsilon$. Therefore, spin flips that do not require large (compared to $T$) activation energy  either pertain to  spins impinging on a monopole and that move the monopole---but monopoles are exponentially few at low $T$---or to collaborative flips of entire loops, which are exponentially unlikely in the size of the loop. Thus, one expects a freeze-in as $T\downarrow0$, in analogy with what is seen in pyrochlores~\cite{snyder2001spin}, but our approach is at equilibrium.

We will therefore intend $T$ as {\em small but not too small} and proceed by assuming that the functional form of an effective theory is quadratic, has the same functional form as the theory above, but that interactions among fluctuations lead to a ``dressing'' in the constants. 

Because correlation lengths diverge at low $T$, from Eq.~(\ref{q2}) we have e.g.\ for charges
\begin{equation}
\langle q^2\rangle  \sim  \chi_0 \xi_{||}^{-2} ~~~~~~~ \text{for} ~T\downarrow 0,
\label{DH}
\end{equation}
which implies a redefinition of constants in the theory at low $T$: $\epsilon, \kappa$ are dressed  as 
\begin{align}
\epsilon \to \epsilon_r(T)& \sim T \chi_0/\langle q^2\rangle \nonumber \\
\kappa \to \kappa_r(T) &\sim T \chi_0/\langle i^2\rangle ~~~~~~~ \text{for} ~T\downarrow 0.
\label{dressing}
\end{align}

Note that this dressing correspond to a quadratic theory that has the correct mean square charge and current already from equipartition. 
Note also that  Eq.~(\ref{DH}) has the form of a Debye screening length for a 2D-Coulomb potential whose coupling constant is proportional to $T$, which is the case for our entropic potential. We have shown elsewhere~\cite{nisoli2020equilibrium} that Eq.~(\ref{DH})  can  follow from a Debye-H\"uckel  approach to the entropic potentials (see also Appendix I).

Note finally that $\chi_0$ must be finite at low $T$ (see also below), and that $\langle q^2\rangle$ is decently approximated by the naive $\overline{q^2}$  
computed by assuming uncorrelated vertices, each with proper multiplicity and Boltzmann weight. At low $T$, $\overline{q^2}\sim (16/3)\exp(-2\epsilon/T)$.
We have thus  
 $\xi_{||}\sim  \exp\left({\frac{\epsilon}{T}}\right)$ for $T\downarrow 0$.
A similar exponential behavior for the correlation length was indeed suggested experimentally by analyzing the pinch points in the structure factor of pryrochlore ice\cite{fennell2009magnetic}. This exponential singularity points to the topological nature of the $T=0$ ice-manifold.  

In conclusion, at low $T$ we can use all the equations of the previous sections, if expressed in terms of $\xi_{||}$, $\xi_{\!\perp}$, where
\begin{equation}
\xi_{||}^2 \sim\begin{dcases}
 \epsilon/T~~~~~~\text{for}~~T/\mu\uparrow \infty \\
\chi_0/\langle q^2 \rangle~~\text{for}~~T/\mu\downarrow 0
\end{dcases}
\label{xinew}
\end{equation}
and similarly
\begin{equation}
\xi_{\!\perp}^2 \sim\begin{dcases}
 \kappa/T~~~~~\text{for}~~T/\kappa\uparrow \infty \\
\chi_0/\langle i^2 \rangle~~\text{for}~~T/\kappa\downarrow 0
\end{dcases},
\label{xinew2}
\end{equation}
and for $\chi_0$, $\chi_0 \sim 1$ for $T\uparrow \infty$ while it remains finite at $T=0$ (We show below that e.g\ for $T=0, \kappa=0$ we have $\chi_0=2$.)

In Appendix A, we further motivate how this choice is mathematically reasonable. 
Note also that no Kosterlitz-Thouless transition of monopole or current unbinding is present~\cite{kosterlitz1973ordering} because the interaction is entropic. Transitions are driven by the interplay between temperature and energy, but here interaction is itself entropic and thus thermal.

\begin{figure}[t!]
\includegraphics[width=.7\columnwidth]{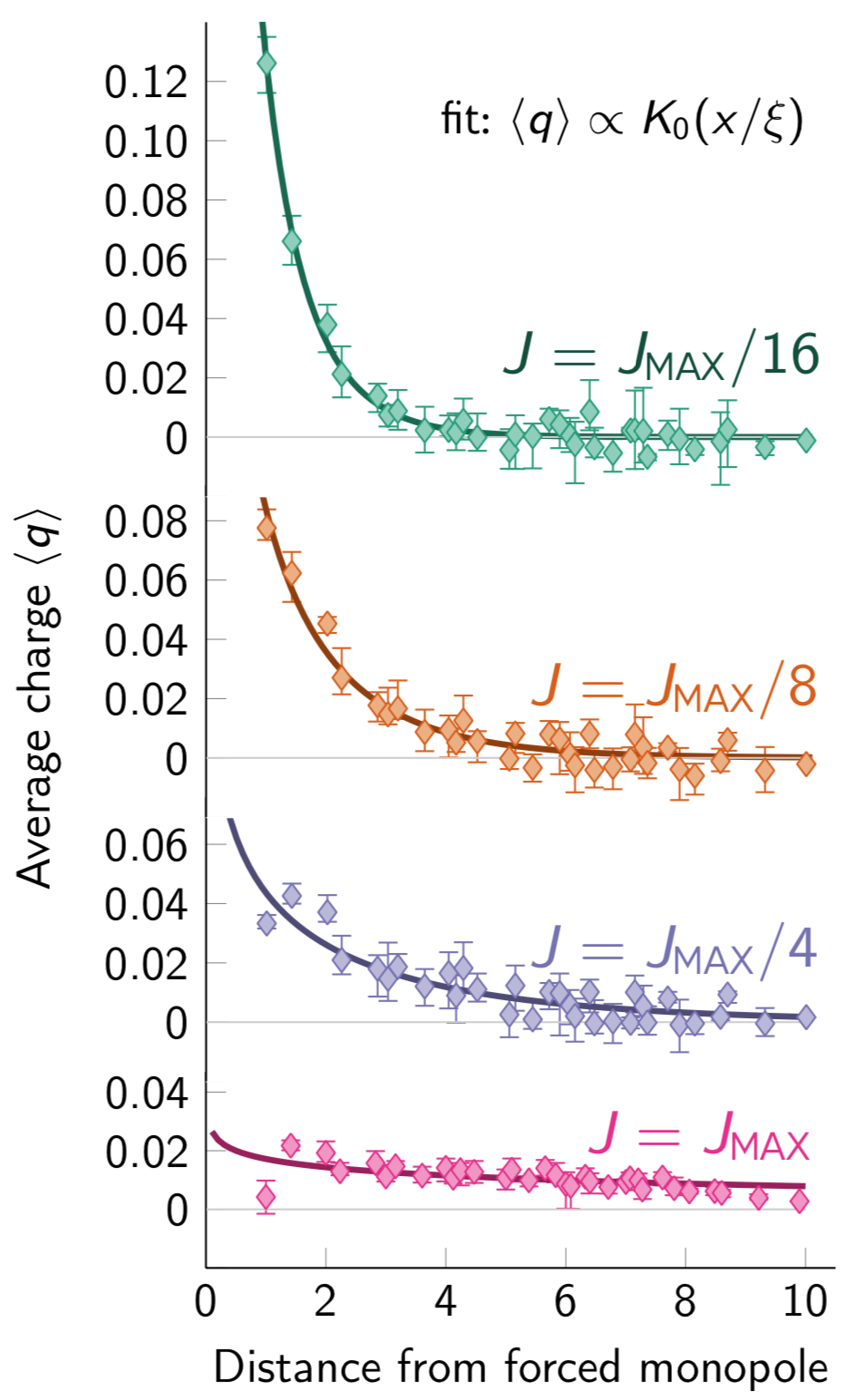}
\caption{Experimental results of entropic screening of a pinned monopole in a square ice realized in a quantum annealer, from ref~\cite{king2021qubit}. Here $J\propto \epsilon/T$ (see reference for details).}.
\label{science}
\end{figure}

\section{Cases}

\subsection{Pure Degenerate Spin Ice ($\kappa=0$)}

In this case, the ground state, aka ice manifold, is the degenerate six-vertex model\cite{lieb1967residual}, described in section \ref{dual}. In the purest form---i.e.\ without the interference of long-range interaction---it was recently realized in a quantum annealer~\cite{king2021qubit}, where the entropic effects were cleanly studied. Nanomagnetic realizations~\cite{Moller2006,perrin2016extensive,farhan2019emergent} imply long range interaction, whose ulterior effects we have described elsewhere~\cite{nisoli2020equilibrium}.

When \hbox{$\kappa=0$}, the functional integration over $\prod_p d i_p$ in Eq.~(\ref{Z2-n}) produces the delta functions $\prod_p \delta(\psi_p)$. Further integration over $\prod_p d \psi_p$ returns for the partition function
\begin{equation}
Z=\int \left[dq\right]  \tilde \Omega[q]  \mathbbm{e}^{-\beta{\cal H}[q]+ \sum_v  q_v V_{q,v}},
\label{Z2-ice}
\end{equation}
where $\tilde \Omega[q]$ is the density of states for the charges, and it is given by
\begin{align}
\tilde \Omega[q,i]=\int [d\phi] \Omega[\phi]\e^{ \sum_v  q_v \ii \phi_v},
\label{Omegatilde-ice}
\end{align}
and is therefore the functional Fourier transform of 
\begin{equation}
\Omega[\phi]=2^{N_e}{\prod_{{\hat{vv'}}}}\cosh \left(-\ii \nabla_{vv'}\phi -\beta \nabla_{pp'}V_{i} +  \beta H_{vv' } \right).
\label{Omega-ice}
\end{equation}

In other words, the currents disappear from the picture, and the entropic potential $-\ii T \psi$ is replaced in the equations by the external potential $V_{i} $ acting on currents. Then, proceeding by the quadratic approximation as before, and using the third of Eq.~(\ref{corr1}), one obtains for the currents correlations 
\begin{equation}
\langle | \tilde i (\vec {k}) |^2\rangle = \chi_{0} \gamma(\vec k)^2,  
\end{equation}
which for small wave vectors reduces to Eq.~(\ref{ii}). 

\subsubsection{Charge Correlations}

More generally, the reader will find that all the equations of the theory developed above apply to the pure ice case by taking $\xi_{\perp}=0$. E.g.,
from Eq.~(\ref{taus}),   $\tilde \chi_{\!\perp}(\vec k)=\chi_0$, and the perpendicular susceptibility in real space is a delta function, consistent with zero correlation length.  

\begin{figure}[t!]
\center
\includegraphics[width=.99\columnwidth]{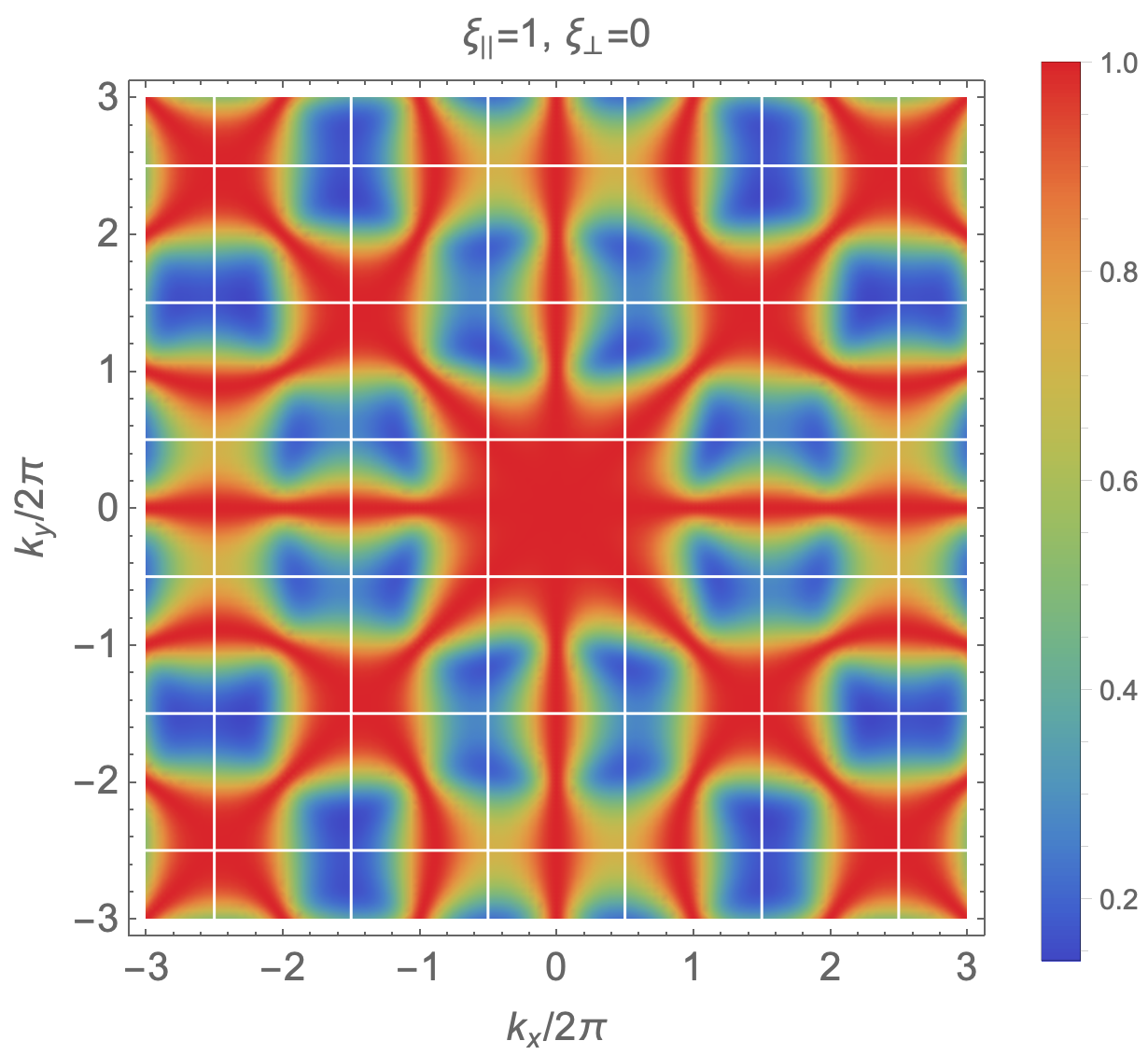}\vspace{3mm}
\includegraphics[width=.99\columnwidth]{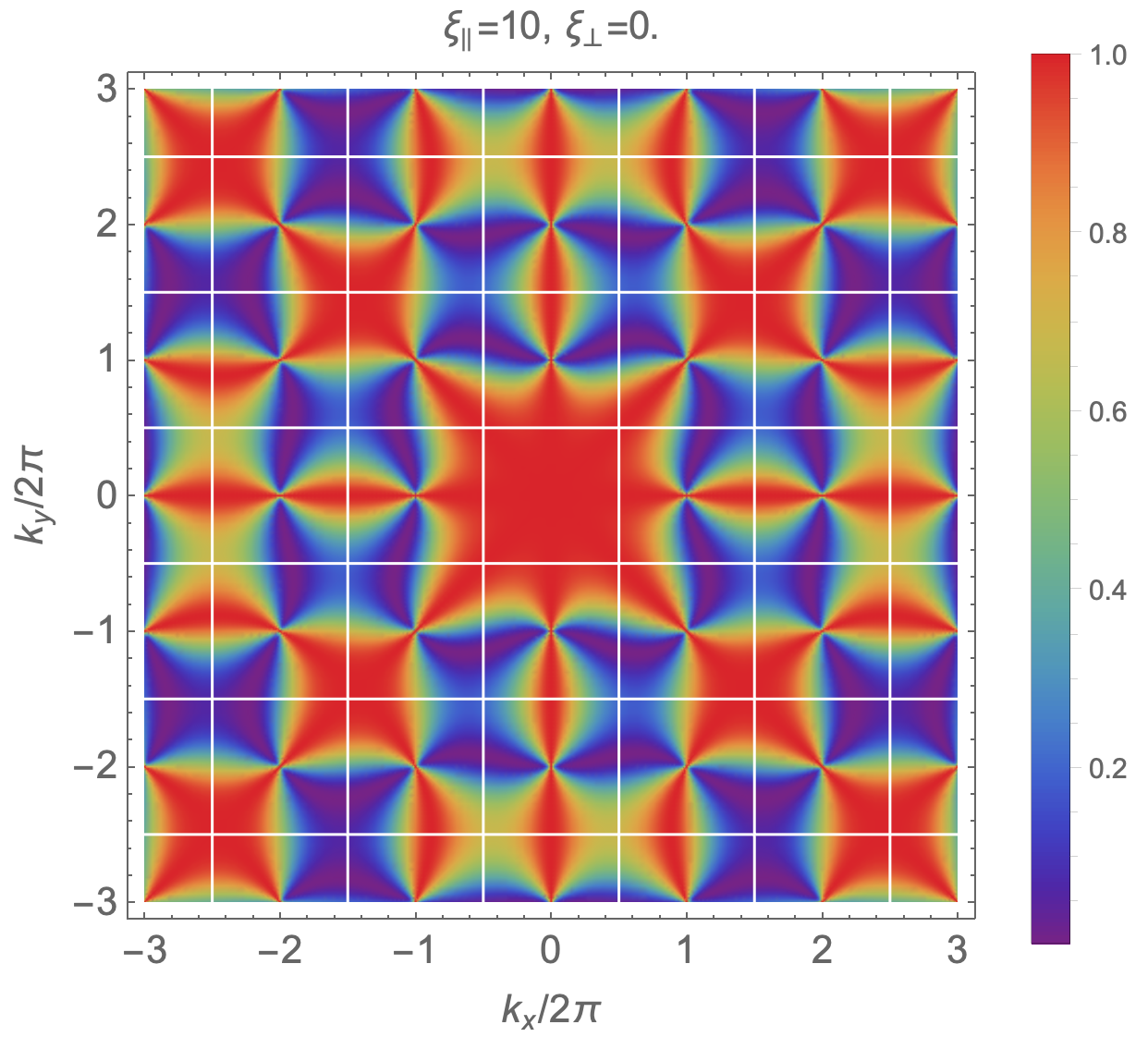} \vspace{-1 mm}

\caption{Structure factors (divided by $\chi_0$) at high (top) and low (middle) temperature plotted from Eq.~(\ref{sigma}), for pure degenerate  square ice. }
\label{IMsf}
\end{figure}

\begin{figure}[t!]
\center

\includegraphics[width=.93\columnwidth]{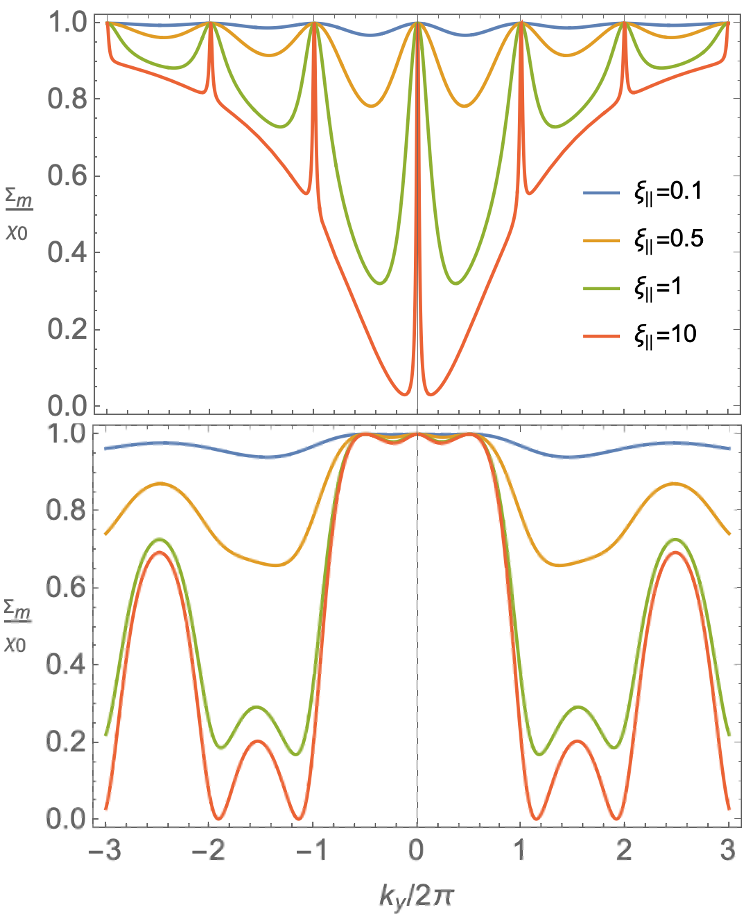}

\caption{Structure factor for pure degenerate  square ice across  the $k_x=2\pi$ line (top)  shows sharpening as the correlation length increases, leading to pinch points, and  across the $k_x=\pi$ line (bottom). }
\label{IMsf2}
\end{figure}

Considering charge correlations and screening, Eq.~(\ref{rhocorr})  can be rewritten as
\begin{equation}
\langle |\tilde q(k)|^2 \rangle= \chi_0 \xi_{||}^{-2}\left(1-\frac{1}{1+\xi_{||}^2 \gamma^2}\right)
\label{pappa}
\end{equation}
where the first term Fourier-transforms to a Kronecker delta, whereas the second term
 implies at large distance the charge correlation
\begin{align}
&\langle q_{v_1} q_{v_2}\rangle = -\frac{\chi_0}{2\pi \xi_{||}^{4}} K_0\left({|v_1-v_2|}/{\xi_{||}}\right). 
\label{K0charge}
\end{align} 
Note that the modified Bessel function $K_0$ is the screened 2D-Coulomb potential. It is exponentially screened, or $K_0(x)\sim \sqrt{\pi/2x}\exp(-x)$ making $\xi_{||}$ the correlation/screening length. In 3D, we would have a screened 3D-Coulomb  or  $\exp(-|v_1-v_2|/\xi_{||})/|v_1-v_2|$.  This result is general: for spin ice on a generic graph, for which Laplacian operators can also be defined, entropic interactions lead to screened Coulomb correlations\cite{nisoli2020concept}. Note  that, from Eq~(\ref{taus}), the kernel of the longitudinal susceptibility functional is also a screened 2D-Coulomb, given by the the modified Bessel function $K_0$ with correlation length $\xi_{||}$.

In artificial realizations it is possible to pin a charge $Q_{\mathrm{pin}}$ in $v_0$.  Then, it is easy to show that the pinned charge generates a charge distribution 
\begin{align}
\langle q_v\rangle=\frac{\langle q_v q_{v_0}\rangle}{\langle  q^2\rangle}Q_{\mathrm{pin}}.
\label{screenp}
\end{align}

   \begin{figure}[t!]
\includegraphics[width=.9\columnwidth]{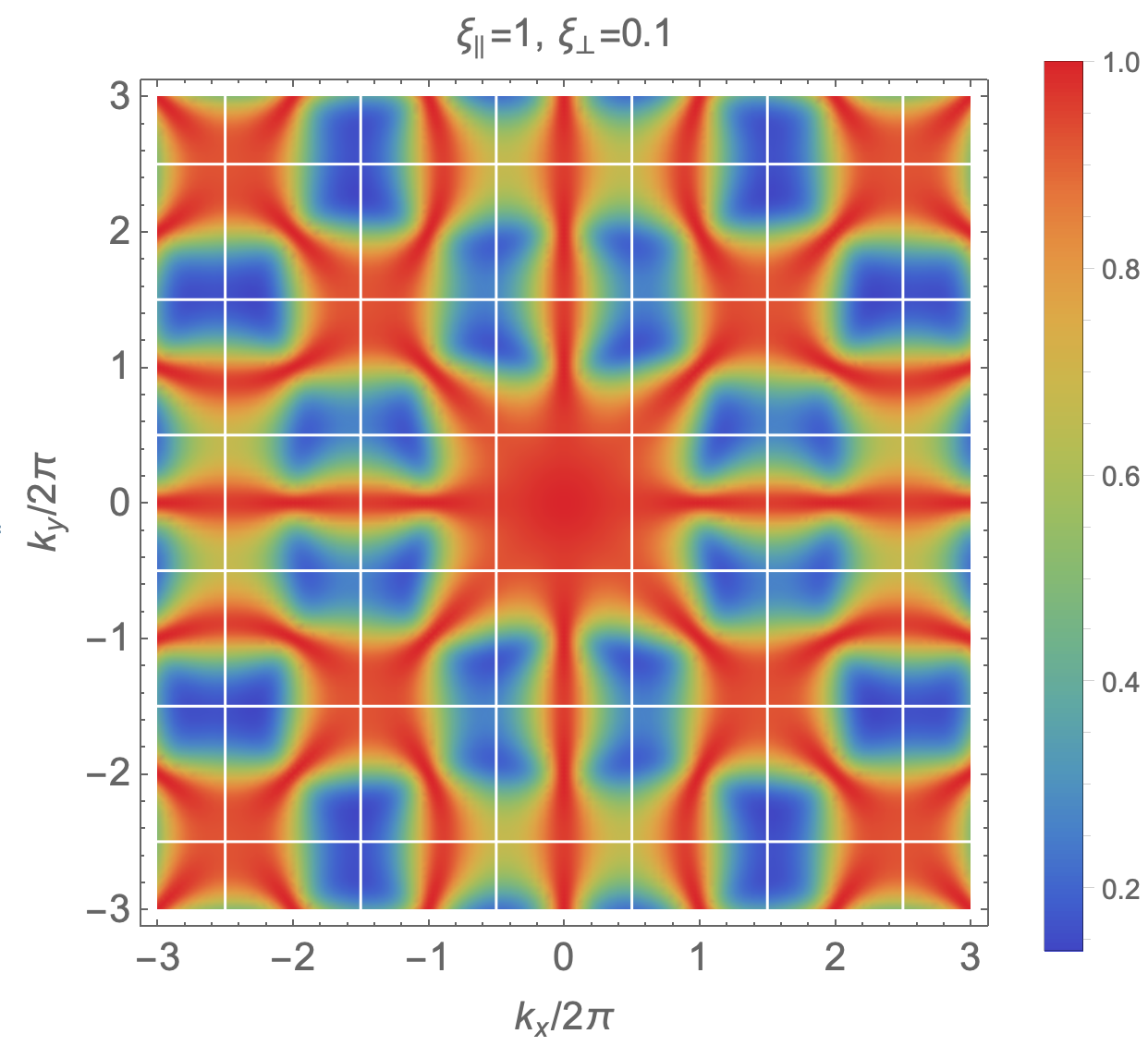}\vspace{0mm}
\includegraphics[width=.9\columnwidth]{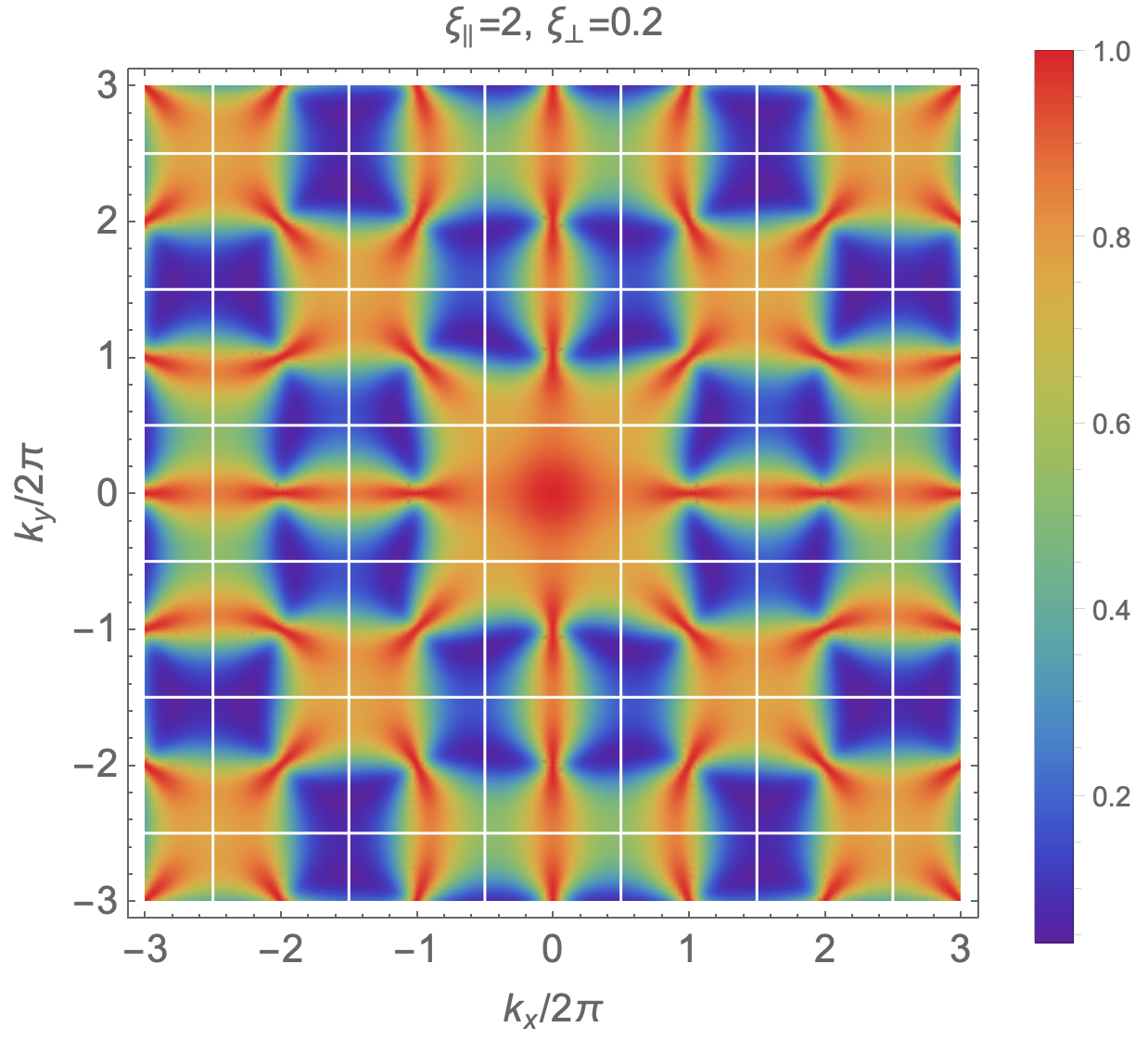}\vspace{0mm}
\includegraphics[width=.9\columnwidth]{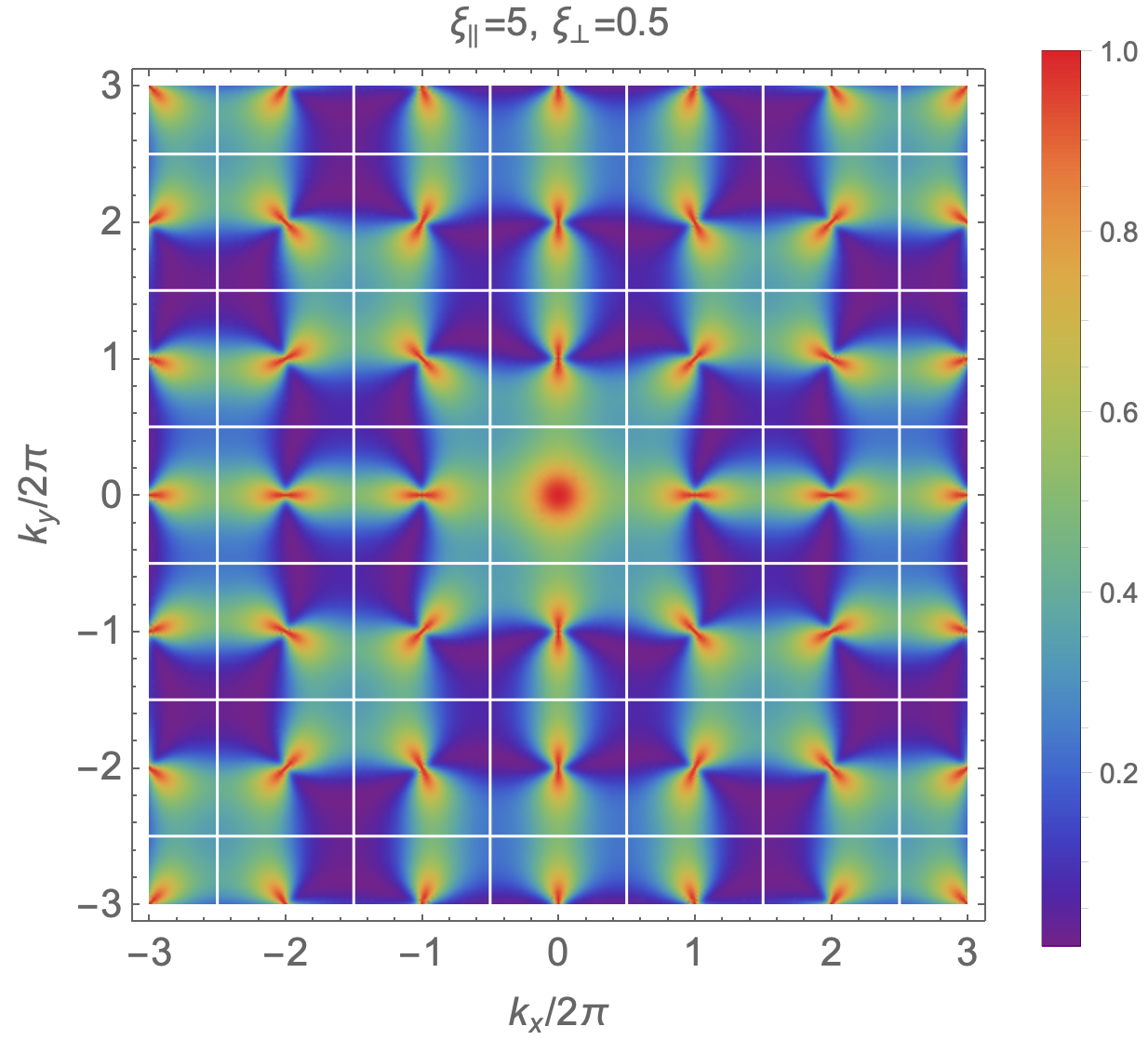}

\caption{Structure factors at different temperatures plotted from Eq.~(\ref{sigma}), for the weak FM square ice. Note the line state emerging, the persistence of pinch points, which are maxima.}
\label{FM}
\end{figure}

\begin{figure}[t!]
\center

\includegraphics[width=.93\columnwidth]{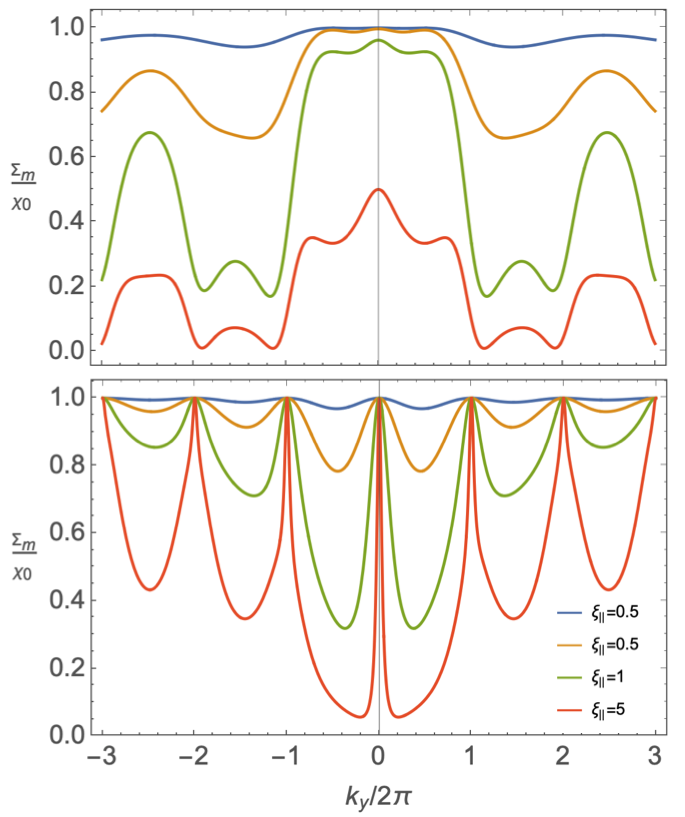}

\caption{Structure factor for  the weak FM square ice  across the $k_x=\pi$ line (top) shows line state features,  and across the $k_x=2\pi$ line (bottom)  shows sharpening as the correlation length increases, leading to pinch points. }
\label{FM2}
\end{figure}

 Remarkably, the screening comes entirely from the entropic interaction. This has been verified in a quantum annealer, where  charges can be pinned~\cite{king2021qubit}. Figure~\ref{science} reports experimental results of screening, which  verify Eq.~(\ref{K0charge}).  At high $\epsilon/T$ the curve becomes flat, consistent with Eqs.~(\ref{pappa}, \ref{K0charge}) as the correlation length exceeds the finite size of the sample~\footnote{ 
The situation becomes considerably more complex in an impure square ice. If monopoles interact also via a {\em real} 3D-Coulomb law, various screening regimes are predicted~\cite{nisoli2020equilibrium}, due to the interplay of the screening length and the Bjerrum length.}.

We  can gain some knowledge of screening at small distances by approximating around the $K=(\pm \pi,\pm \pi)$ points of the BZ. There, $\gamma(k)^2$ is maximum and $\gamma(\vec k+K)^2=8-k^2$. This leads to a screening function
\begin{equation}
\langle |\tilde q(K+k)|^2\rangle\simeq\xi_{||}^{-2}\left(1-\frac{\xi_{||}^{-2}}{\xi_p^{-2}- k^2}\right)
\end{equation}
where $\xi_p^2 = \xi_{||}^2/(1+8 \xi_{||}^2)$. Then for small $\Delta v=v-v'$ the charge correlation (or equivalently screening) has a sign alternation with the Manhattan distance on the graph and an envelope function $E(|v-v'|/\xi_p )$ of periodicity $\xi_p/2\pi$,  of the form
\begin{align}
\langle q_v q_{v'}\rangle = (-1)^{\Delta v_x+ \Delta v_y}E(|\Delta v|/\xi_p) +\text{constant}.
\end{align}
Such sign alternation with the Manhattan distance for screening at short distance was also verified in  experiments~\cite{king2021qubit} on a quantum annealer\footnote{Note, however, that monopoles are characterized not only by a charge, but also by a net magnetic moment, which was not taken into account in our formalism. Therefore, at short distances the screening would be anisotropic, just as monopoles are. In a future work we will incorporate that degree of freedom to study at short length or equivalently around the $K$ points of the BZ.}.

\subsubsection{Spin Correlations}

Considering now the spin correlations, Eq.~(\ref{spincorr}) particularizes now to
\begin{align}
\chi_0^{-1}\langle \tilde S^*_{\alpha}(\vec k) \tilde S_{\alpha'}(\vec k)\rangle &= \delta_{\alpha\alpha'} -  \hat \gamma_{\alpha} \hat \gamma_{\alpha'} \left(1- \tilde \chi_{||}/\chi_0\right) \nonumber \\
 &=\delta_{\alpha\alpha'} -  \hat \gamma_{\alpha} \hat \gamma_{\alpha'} \xi_{ ||}^{2} |\tilde q(k)|^2 \nonumber \\
 &=\delta_{\alpha\alpha'} - \frac{ \gamma_{\alpha} \gamma_{\alpha'}}{\xi_{ ||}^{-2} + \gamma^2}.
\label{spincorrIM}
\end{align}

In real space  at  distances larger than lattice discretization we have
\begin{equation}
\chi_0^{-1} \langle S_{\alpha}(\vec x)  S_{\alpha'}(\vec 0)\rangle\simeq - 2\pi \partial^2_{\alpha \alpha'} K_0(|\vec x|/\xi_{||}),
\label{spincorrIMreal}
\end{equation}
which is  {\em not} algebraic at non-zero temperature. At long wavelength, the spin correlations in real space in the pure ice manifold are thus the  kernel of a {\it screened}, 2D dipolar interaction, i.e.\ obtained as partial derivatives of the screened 2D-Coulomb interaction ($K_0$), thus generalizing the heuristic Eq.~(\ref{heighcorr2}). 

Spin correlations become algebraic in the (unrealistic)  $T=0$ manifold, where $\xi_{||}^{-1}=0$. Then,  for the pure ice manifold we have the familiar transverse form
\begin{align}
\chi_0^{-1} \langle \tilde S^*_{\alpha}(\vec k) \tilde S_{\alpha'}(\vec k)\rangle_{\text{IM}}=  {^{\!\perp} \! \hat \gamma_{\alpha}} \! \!{^{\!\perp} \! \hat \gamma_{\alpha'}}=\delta_{\alpha \alpha'} - \frac{\vec \gamma_{\alpha} \vec \gamma_{\alpha'}}{\gamma^2}.
\label{SSice}
\end{align}
That is because,  the longitudinal susceptibility $\tilde \chi_{||}$ goes to zero in the ice manifold, as it is associated to monopoles, which disappear. Thus only the transverse part of Eq.~(\ref{spincorr}) remains. Indeed, temperature adds a longitudinal part to the spin correlations, such that the difference between correlations with and without temperature is
\begin{equation}
\langle \tilde S^*_{\alpha}(\vec k) \tilde S_{\alpha'}(\vec k)\rangle-\langle \tilde S^*_{\alpha}(\vec k) \tilde S_{\alpha'}(\vec k)\rangle_{\text{IM}}= \chi_0 \hat \gamma_{\alpha} \hat \gamma_{\alpha'} \tilde \chi_{||}.
\end{equation}

In Fig.~\ref{IMsf} we plot the  structure factor $\Sigma_m(\vec k)$ in units of $\chi_0$  from Eq.~(\ref{spincorrIM}), demonstrating the formation of pinch points as $T$ is reduced, shown in Fig.~\ref{IMsf2}, top. Note that they compare very  favorably with  structure factors obtained experimentally in a quantum annealer (Fig.~2 of ref~\cite{king2021qubit}) in the case of a line state, as well as from simulations~\cite{rougemaille2021magnetic,brunn2021signatures}. Interestingly, Fig.~\ref{IMsf2} bottom show light bumps in the $K$ points of the BZ, that are also see, though more pronounced, in experimental and numerical results, and correspond to small AFM domains of t-I in the Ice Manifold, whose net AFM staggered order parameter is nonetheless zero (no symmetry breaking).

Finally, spin correlations allow us to say something about $\chi_0$. 
Because  spins have values \hbox{$S_{\alpha}=\pm1$}, then \hbox{$1=\langle S^2_{\alpha} \rangle= N_v^{-1} \langle \sum_e S^2_{\alpha,e} \rangle=\int_{\text{BZ}}\langle |\tilde S_{\alpha}(\vec k)|^2\rangle_{\text{IM}}\frac{d^2k}{(2\pi)^2}$}, and summing over \hbox{$\alpha=x,y$}, from Eq.~(\ref{SSice}) we obtain 
\begin{equation}
\chi_0=2~~~~\text{for}~~~~T=0.
\end{equation}

Proceeding in the same way, but using the second line of Eq.~(\ref{spincorrIM}) we obtain for generic temperature
\begin{equation}
\chi_0= 1+\frac{1}{2} \xi_{||}^2 \langle q^2\rangle,
\label{seth}
\end{equation}
which relates the uniform susceptibility to the correlation length and the mean square charge, at all temperatures. Then from Eqs.~(\ref{xinew}, \ref{seth}) we find again $\chi_0 \to 2$ for $T\downarrow 0$ and $\chi_0 \to 1$ fo $T\uparrow +\infty$.

\begin{figure}[t!]
\includegraphics[width=.99\columnwidth]{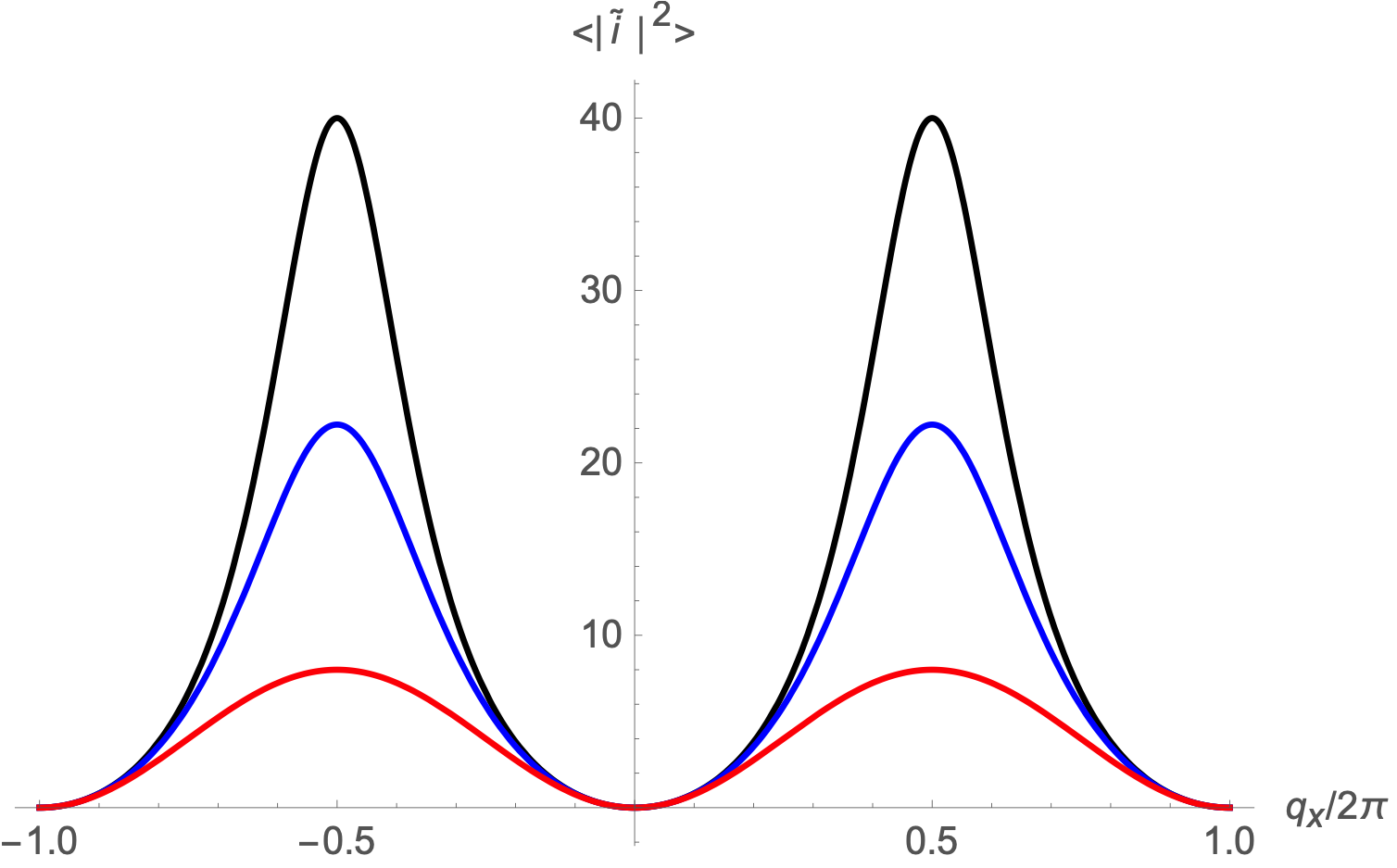}

\caption{Plot of $\langle | \tilde  i (\vec {k}) |^2\rangle\chi_0^{-1}$ along the line $k_x=k_y$ for different negative values of $\kappa/T$ ($\kappa/T=0$ in red, $\kappa/T=0.08$ in blue, $\kappa/T=0.1$ in black) showing the growing divergence at the $K$ points $ k_x=k_y=\pi$.}
\label{AFM2}
\end{figure}

\begin{figure}[t!]
\includegraphics[width=.99\columnwidth]{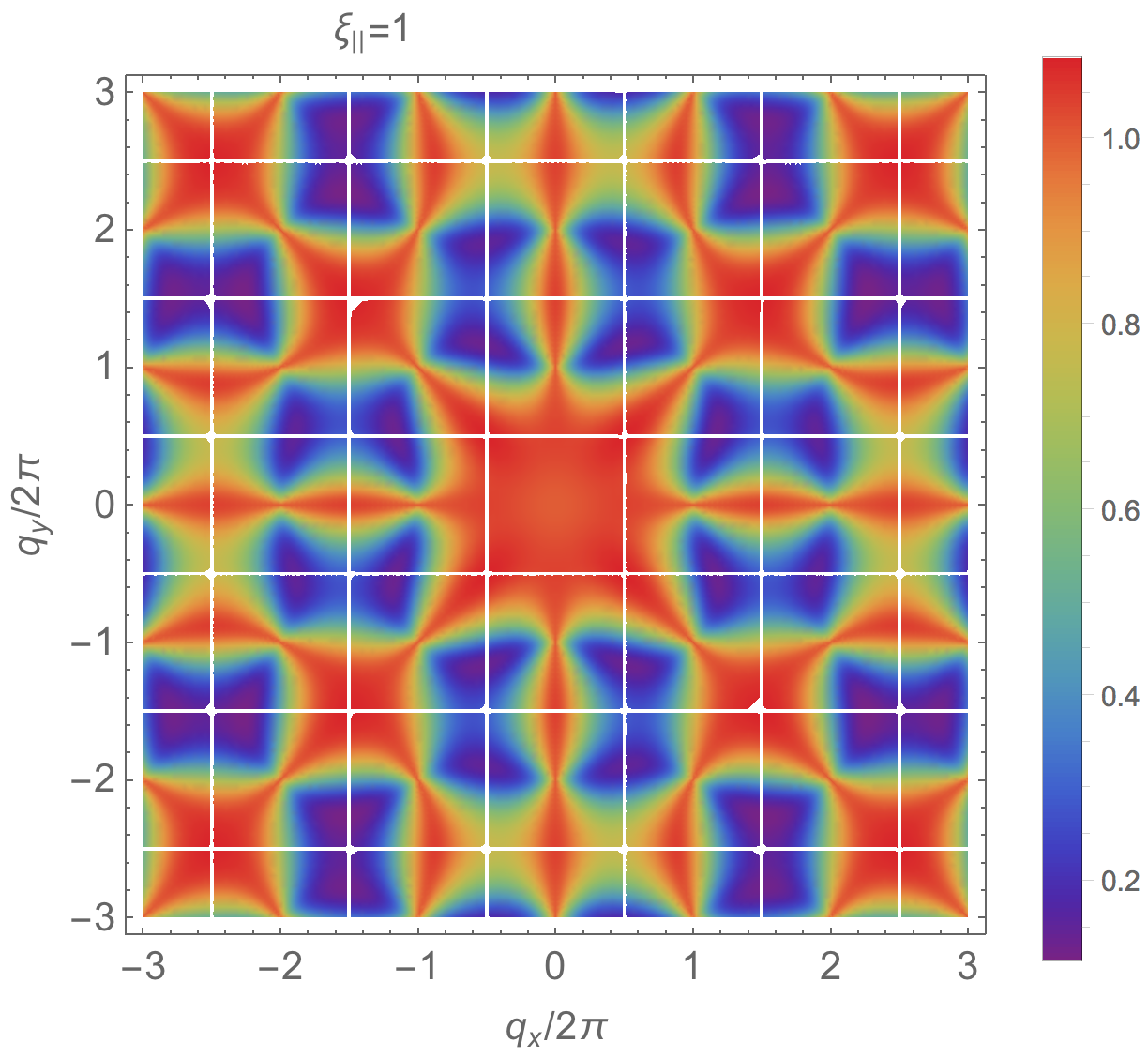}
\includegraphics[width=.99\columnwidth]{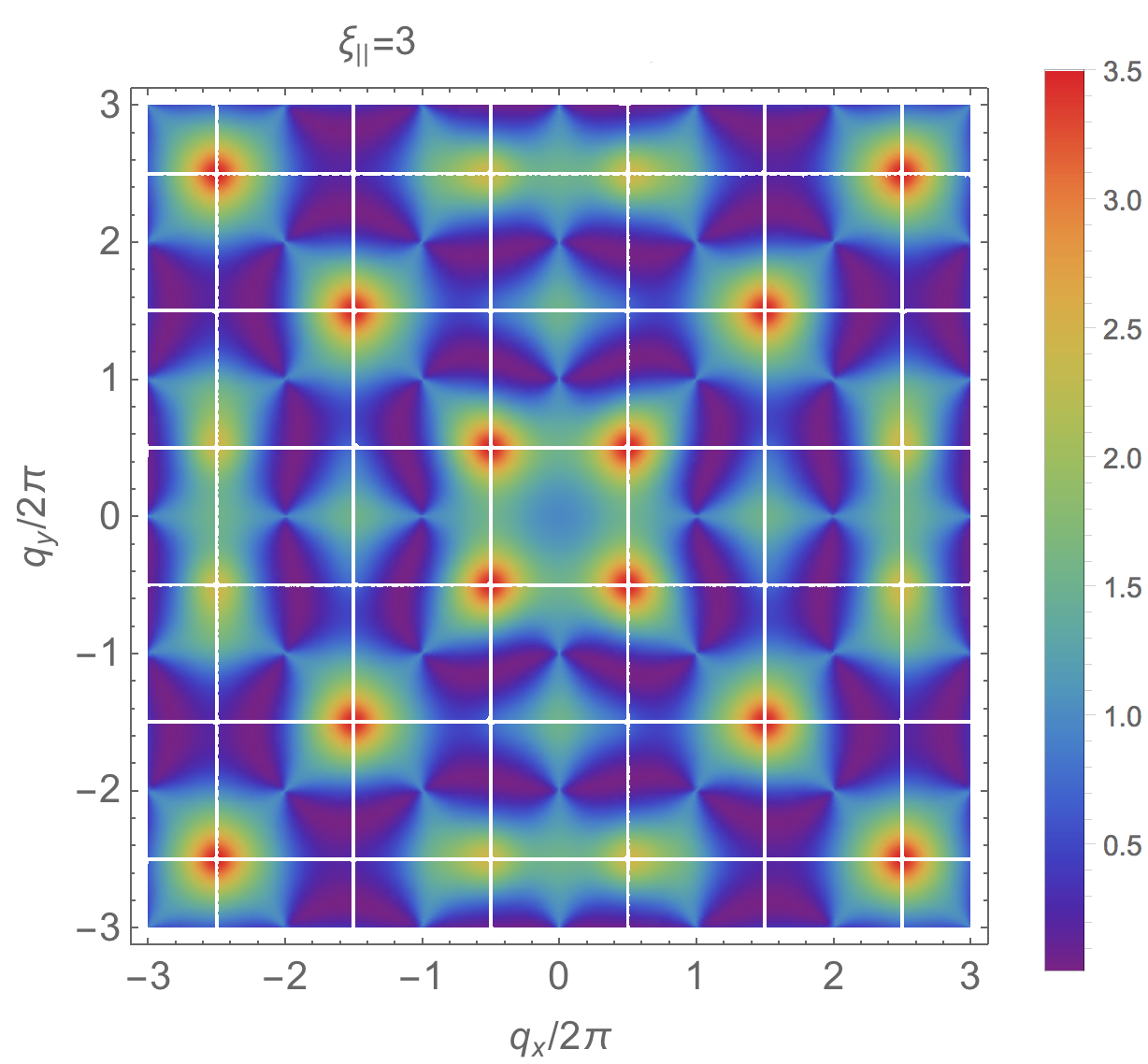}
\caption{Structure factors at different temperatures plotted from Eq.~(\ref{sigma}), for the weakly AFM square ice  when $\xi_{\perp}=0.1  \xi_{||}  \ii$. Note the formation of peaks at the $K$ points of the Brillouin zone indicating FM ordering.}
\label{AFM}
\end{figure}

\subsection{Weak Ferromagnetic Case ($\kappa \simeq 0^+$)}

This case  has an ordered, fourfold, ferromagnetic ground state, corresponding to the four possible full polarizations of the system. We are interested here in small $\kappa/\epsilon$ (hence ``weak"), at temperatures above the ordering, whose scale is set by $\kappa$, but below the crossover into the ice manifold, whose scale is set by $\epsilon$.  While this case does not describe at low temperature  square ice with heigh offset exceeding the critical offset~\cite{Moller2006,perrin2016extensive}, which instead leads to a disordered, albeit subextensively so, line liquid,  it is likely a good approximation for it at intermediate temperatures. 

Because of the gauge-free duality, everything said above for charge correlations applies here both to charges and currents. Thus, on top of the previous equations, there are also screened correlations for currents as
\begin{align}
&\langle i_p i_{0}\rangle =  -\frac{1}{2\pi \xi_{\perp}^2} K_0\left({|p|}/{ \xi_{\! \perp}}\right). 
\label{K0}
\end{align} 
In particular, a pinned current generates a screening from other currents, similar to the screening of a charge.

While the considerations in the previous subsections on charge correlations still apply, the spin correlations are now different and are given by the full Eq.~(\ref{spincorr}).
The structure factor (Figz.~\ref{FM}, \ref{FM2}) remains reminiscent of  spin ice. As temperature is reduced it also shows the features of the line state (see Fig.~2e in ref~\cite{rougemaille2021magnetic} and Fig~2 in ref~\cite{king2021qubit}). However, it also show a maximum at the center of the BZ that signals incipient ferromagnetic ordering, absent in the standard line state, where instead it is replaced by  maxima on the $k_x, k_y$ axes. 

These features   compare very  favorably with those in structure factors obtained experimentally in a quantum annealer (Fig.~2 of ref~\cite{king2021qubit}) in the case of a line state, as well as from simulations in refs~\cite{rougemaille2021magnetic,brunn2021signatures}. 

\subsection{Weak Antiferromagnetic Case ($\kappa \simeq 0^-$)}
\label{kneg}

When $k<0$,  ${\cal H}_2$ in Eq.~(\ref{f2rho}) is not bounded from below when $T\le T_c^{\text{afm}}= 8 |\kappa|$ (because $\gamma^2$ reaches its maximum $\gamma^2=8$ on the $K$ corners of the BZ). This  merely testifies to the expected ordering criticality in the AFM case. Currents are promoted by a negative $\kappa$ and the AFM state is an ice rule state that maximizes currents, made of a tessellation of t-I vertices. There is a second order phase transition to such ordered state. Because we have employed a mean field theory,
 $T_c^{\text{afm}}$ is not the actual  critical temperature, but merely a useful parameter in the context of our framework.

 For $T>T_c^{\text{afm}}$, $\tilde \chi_{\!\perp}(k)$ has a maximum on the $K$ corners of the BZ. 
Thus, we expand around $K$, $\gamma (K +\vec k)^2 \simeq 8- k^2$ in  $\tilde \chi_{\!\perp}(k)$ and from Eq~(\ref{rhocorr}) we obtain for large $|p|$
\begin{align}
&\langle I_p I_{0}\rangle \simeq \frac{(T/\kappa)^2}{2\pi} (-1)^{p_x+p_y } K_0\left({|p|}/{\xi_{\text{afm}}}\right), 
\label{K0AFM}
\end{align} 
which expectedly alternates sign on adiacent plaquettes. 

The AFM correlation length is given by
\begin{equation}
\xi_{\text{afm}}^2=|\kappa|/(T-T_c^{\text{afm}})
\end{equation} 
 and is a measure of the size of the antiferromagnetic domains. It diverges at the ordering criticality, though with the wrong critical exponent. Since one expects a 2D Ising transition\cite{Wu1969,anghinolfi2015thermodynamic}, the exponent should be $1$. Here it is $1/2$, or the mean-field exponent, as we have employed a quadratic mean-field approximation.
 
Because $\kappa/\epsilon$ is small, $T^{\text{afm}}$ is much smaller than the crossover temperature for the ice regime (typically $\sim2\epsilon$). Thus, above $T_c^{\text{afm}}$  the divergence-free and divergence-full fields behave independently and features of the essentially transversal IM structure factor, such as pinch points and charge correlations, are still present. 
We plot the structure factor in Fig.~\ref{AFM}. Note there the. growing maxima at the $K$ points of the BZ, corresponding to AFM ordering. 

Again, these features   compare very  favorably with those in structure factors obtained experimentally in a quantum annealer (Fig.~2 of ref~\cite{king2021qubit}) in the case of a line state, as well as from simulations in refs~\cite{rougemaille2021magnetic,brunn2021signatures}.

\section{Na\"ive Kinetics}

The formalism above can be used for general considerations about kinetics on time scales comparable to the relaxation times or shorter~\cite{chaikin1995principles}. We anticipate here the main deductions, leaving a more in-depth study to future work on ``slow'' electrodynamics of spin ice. 

\subsection{General Considerations}
Unlike equilibrium thermodynamics, which merely concerns itself with samplings of the phase space, dynamics is generally system-specific. Therefore, many different models of kinetics  could be introduced, in or out of equilibrium, closer to the constitutive properties of a specific material or more general. In this subsection, we consider notions of ``kinematics'' that  hold true in any such model. 

As noted already~\cite{ryzhkin2005magnetic}, simple considerations should readily convince  that the flow density vector~\footnote{We use here ``flow density vector'' to denote what is more generally called ``current density vector'', to avoid confusion with the notion of currents $i$ previously introduced.} for the charges, in or out of equilibrium, is always given by
\begin{equation}
\vec J^{q}=\dot{\vec S},
\label{trivial}
\end{equation}
in perfect analogy with the relation between electrical current and the dielectric polarization vector in standard electrodynamics. 

Reasoning in the long wavelength limit, and taking the divergence we obtain the conservation equation for the charge
\begin{equation}
\dot q=-\vec \nabla \cdot \vec J^{q},
\label{qcons}
\end{equation}
which is merely the derivative of Eq.~(\ref{tito}).
Similarly, taking instead the curl, we have 
\begin{equation}
\dot{i}=\hat e_3 \cdot \vec \nabla \wedge \vec J^{q}= -\vec \nabla \cdot \vec J^{i}
\label{icons}
\end{equation}
with
\begin{equation}
\vec J^{i}=\hat e_3 \wedge \vec J^{q}.
\label{icons}
\end{equation}
We found that the flow density vectors for charges $q$ and currents $i$ are orthogonal, as expected by gauge-free duality. Thus a flow of monopoles always implies a flow of currents, and vice versa. 

Even a distracted look at the physical system should convince that there cannot be steady states of charge flow. After the application of a uniform field, the spins reorient. This can be interpreted as flow of monopoles and currents. 
At equilibrium all the spins have reached the new configuration, the system is static and there is no more flow. Direct current is therefore only possible during that relaxation, and it is not steady~\footnote{Depending on the system, the aging might be slow rather than exponential~\cite{castelnovo2010thermal,levis2013defects}, in which case tiny currents might be detected after a long time.}.


\subsection{Relaxation Dynamics}

As mentioned, a simple relaxation dynamics at equilibrium would not apply to real materials, as it is clear both experimentally~\cite{bovo2013brownian,dusad2019magnetic,goryca2021field} and conceptually~\cite{castelnovo2010thermal,levis2013defects,nisoli2020colored}. Depending on the system, however, deviations can be limited to kinetics much faster than relaxation~\cite{nisoli2020colored}.

A fundamental aspect is the choice of the degrees of freedom that relax. In our system it seems reasonable to choose the spins as ``real variables'', and not charges or currents, nor, as done by e.g. C. Henley~\cite{henley1997relaxation} in the context of the similar dimer model, the height functions.  
Thus we proceed with the relaxation equation for the spins $S$ which we write as
\begin{equation}
\tau_0 \dot{\vec S}=-\beta\frac{\delta {\cal F}[\vec S]}{\delta \vec S}
\label{relax}
\end{equation}
where $\tau_0$ is a characteristic time of the kinetics, and changes with temperature. 

\subsubsection{Dynamic Susceptibilities, Conductivities, Dispersion}

From the Eqs.~(\ref{relax}, \ref{FS})   we obtain for the longitudinal and transverse spin components the equations of motion in reciprocal space.
\begin{align}
\tau_0 \dot{\tilde S}_{||,\perp}(\vec k)= -\tilde \chi_{||,\perp}(\vec k)^{-1} \tilde S_{||,\perp}(\vec k) + \beta \tilde H_{||,\perp}. 
\label{kinS}
\end{align}
%
Thus, the longitudinal and transverse relaxation times  are
\begin{equation}
\tau_{||,\perp} (\vec k)=\tau_0 \tilde \chi_{||,\perp}(\vec k).
\label{taure}
\end{equation}
and thus the relaxation frequencies (from $\nu=1/\tau$) obey the dispersion relations
\begin{equation}
\nu_{||,\perp} (\vec k)= \frac{1}{\chi_0\tau_0} + D_{||,\perp} \gamma (\vec k)^2,
\label{nure}
\end{equation}
where $D_{||}, D_{\!\perp}$ are diffusion constants (see below) given by 
 \begin{equation}
 D_{||,\perp}=\frac{\xi^2_{||,\perp}}{\tau_0 \chi_0}.
 \label{D}
 \end{equation}
[Also, from Eq.~(\ref{D}, \ref{taure}) one finds Eq. (7.5) of ref~\cite{bramwell2012generalized}.]

From Eq.~(\ref{taus}) we see that when $\epsilon>0, \kappa>0$, the relaxation time is minimal at the $K$ points of the Brillouin zone, which corresponds to  processes that involve few, neighboring spins, e.g.\ the creation/annihilation of a monopole pair. When $\kappa=0$ the transverse relaxation time $\tau_{\!\perp}(\vec k)$ is flat and equal to $\tau_0$ because processes concerning currents do not change the energy. 

From Eq.~(\ref{kinS}) we find the dynamic susceptibilities (defined for dimensional convenience as $\tilde S=\tilde \chi \beta \tilde H$) are
\begin{align}
\tilde \chi_{||,\perp}(\omega,\vec k)=\frac{\chi_{||,\perp}(\vec k)}{1-\ii \omega \tau_0 \chi_{||,\perp} (\vec k)}
\label{chidyn}
\end{align}
as already appreciated via other methods in pyrochlores spin ice~\cite{bramwell2012generalized}.

The dynamic susceptibilities in Eq.~(\ref{chidyn}) are typical of an exponential relaxation to equilibrium~\cite{topping2018ac}.  They are wrong at high frequency. Indeed, Eq.~(\ref{chidyn})  and the fluctuation-dissipation theorem imply that the power spectrum is a Lorentzian, and correspond to a brown noise, i.e.\ scaling as $1/\omega^2$ at large $\omega$. Instead,  recent experiments report a ``color'' of the noise that depends on temperature and varies between brown and pink~\cite{dusad2019magnetic,goryca2021field}, suggesting instead a generalized Debye function~\cite{cole1941dispersion}, further underscoring the expected limits at frequencies of a mean field relaxation kinetics\cite{nisoli2020colored}. 

From the first of Eq.~(\ref{kinS}), and remembering $\tilde q=\gamma \tilde S_{||}$,  $\tilde i=\gamma \tilde S_{\!\perp}$ we obtain the relaxation equation for the charge as
\begin{align}
\tau_0 \dot{\tilde q}(\vec k)= -\left[1 + \xi_{||}^2 \gamma(\vec k)^2 \right]\tilde q(\vec k) -\ii \beta \vec \gamma \cdot \tilde H \nonumber \\
\tau_0 \dot{\tilde i}(\vec k)= -\left[1 + \xi_{\!\perp}^2 \gamma(\vec k)^2 \right]\tilde q(\vec k) -\ii \beta ^{\perp\!}\vec \gamma \cdot \tilde H,
\label{kinQ}
\end{align}

Finally, from Eqs.~(\ref{trivial}-\ref{chidyn}) we have, for the longitudinal and transverse conductivity, defined as $J^q=\sigma H$, the expression
\begin{align}
\tilde \sigma_{||,\perp}(\omega,\vec k)=\frac{1}{T\tau_0}\frac{-\ii \omega \tau_{||,\perp}(\vec k)}{1-\ii \omega \tau_{||,\perp}(\vec k)},
\label{sigmadyn}
\end{align}
which are zero at $\omega=0$, as it should be: there are no direct currents. 

\subsubsection{Real Space Picture}
 Equations~(\ref{kinQ}) become,
 in real space at length scales larger than the lattice constant,  diffusion-relaxation equations
\begin{align}
\dot{ q}= D_{||}\Delta q + \left( \beta \nu q_{\text{ext}}-q \chi_0^{-1}  \right) \tau_0^{-1} \nonumber \\
\dot{ i}= D_{\!\perp}\Delta i + \left( \beta \nu i_{\text{ext}}-i \chi_0^{-1} \right)\tau_0^{-1}. 
\label{kinQr}
\end{align}

The first term in Eqs.~(\ref{kinQr}) accounts for the diffusion of charges or currents. The diffusion flow density vectors are  
\begin{align}
\vec J^{q,\text{diff}}=- D_{||} \vec \nabla q \nonumber \\
\vec J^{i,\text{diff}}=- D_{\! \perp} \vec \nabla i,
\label{Jqdiff}
\end{align}
for charges and currents respectively.

The second term in Eqs.~(\ref{kinQr}), in absence of external charges, accounts for  pair annihilation of charges and relaxes exponentially the charge to the equilibrium value $\langle q \rangle=0$. 
In presence of external charges the system relaxes to the screening Equations~(\ref{screen}).

A similar diffusion-relaxation equation can be written for $\vec S$. From Eq.~(\ref{kinS}) we have
\begin{align}
 \dot{S_{\alpha}}&= D_{||} \nabla_{\alpha}  \vec \nabla \cdot \vec S + D_{\!\perp} \!\left[\vec \nabla \wedge \left( \vec \nabla \wedge \vec S\right) \right]_{\alpha} \!\! \nonumber \\
 & + \left(\beta H_{\alpha}-S_{\alpha} \chi_0 ^{-1}\right)\tau_0^{-1}
 \label{kinSr}
\end{align}
which can also be written as
\begin{align}
 \dot{S_{\alpha}}&=  -\nabla^{\alpha'} J^S_{{\alpha'} {\alpha}}+ \left(\beta H_{\alpha}-S_{\alpha} \chi_0 ^{-1}\right)\tau_0^{-1} 
\label{kinSr2}
\end{align}
in terms of the flow density tensor for magnetization 
\begin{align}
J^S_{{\alpha'} {\alpha}}= -D_{||}\nabla_{\alpha} S_{\alpha'} -D_{\! \perp}\left(\nabla_{\alpha}S_{\alpha'}-\nabla_{\alpha'}S_{\alpha}\right).
\label{Js}
\end{align}
However magnetization is not conserved, but because of the second term $-S_{\alpha}$, it decays exponentially to the equilibrium value $\langle S_{\alpha} \rangle=0$ imposed by the $Z_2$ symmetry of the problem (if $\vec H=\vec 0$).

Equation~(\ref{kinSr}) can also be written in terms of charges and currents
\begin{align}
 \dot{\vec S}&= -D_{||} \vec  \nabla q -D_{\!\perp} \hat e_3 \wedge \vec \nabla i + \left(\beta H_{\alpha}-S_{\alpha} \chi_0 ^{-1}\right)\tau_0^{-1},
\label{kinSr2}
\end{align}
from which Eqs.~(\ref{kinQr}) can be deduced directly by taking the divergence and the curl.
Then, in Eq.~(\ref{kinSr2}) we recognize the diffusion flow density vectors of Eq.~(\ref{Jqdiff}). We can therefore write for the {\em total} flow density vector of the charge
\begin{equation}
\vec J^q= \vec J^{q,\text{diff}} + \hat e_3 \wedge \vec J^{i,\text{diff}} + \left(\beta H_{\alpha}-S_{\alpha} \chi_0 ^{-1}\right)\tau_0^{-1}.
\label{Jtot}
\end{equation}
The second term is clearly divergence-free. From Eq.~(\ref{icons}) the flow density vector of the currents is obtained as its rotation.

The third term in Eqs.~(\ref{kinSr2},\ref{Jtot}) is a drift term to which magnetization is subtracted. Something similar was previously found in pyrochlore spin icevia considerations of chemical physics of electrolytes~\cite{ryzhkin2005magnetic,bramwell2012generalized}. At equilibrium, for a constant uniform field $\vec H$,  that term is zero, currents and charges are uniform, and thus there is no longer monopole flow. Again, direct current of monopoles is only possible during relaxation. 

When $\kappa=0$, $D_{\perp}=0$ and the picture simplifies. Currents no longer diffuse, but relax uniformly. However  from Eq.~(\ref{icons}) there is a divergence free flow of currents, if monopoles are flowing.

\section{Conclusions}

We have illustrated the duality between charges and currents in pure spin ice, that is square spin ice coupled only at the vertex level. We have built on it a field theory where elementary currents and monopoles are the degrees of freedom, while spins are subsumed  into entropic interactions. In pure spin ice, this leads to a 2D electrodynamics formalism where 2D-Coulomb interactions are entropic.  

Within this framework, we have deduced free energies, static and dynamic susceptibilities, relaxation times, form factors, structure factors, for the three cases of: degenerate spin ice, line state, antiferromagnetic square ice. Our purely analytical results compare  well with a wealth of experimental and numerical data. They generalize to thermal states previous heuristic approaches at zero $T$ variously developed in the context of the dimer model, and based on the height function formalism. They also accord well with chemical physics approaches to 3D pyrochlores.

\begin{appendices}

\section{Conceptualizing the Dressing at Low $T$}

To understand the mathematical origin of the dressing, consider that  $\Omega[\phi,\psi]$ is periodic in the gradient of the entropic fields. Such periodicity comes from the sum over the spin ensemble of the Fourier transform of Dirac deltas. The latter enforce the discrete nature of charges and currents.
Thus, $\Omega[\phi,\psi]$ has two roles: one is to convey the entropic interaction, and the other is to preserve the information that charges are discrete. In our high temperature limit, however, we had taken $\nabla \phi, \nabla \psi$  small, thus losing the periodicity needed to constrain the magnitude of charges. In this scenario, the low temperature dressing [Eqs.~(\ref{dressing})]    within an effective quadratic theory takes care of that constraint already at the level of equipartition, while maintaining a formalism of continuum charge distribution.

To separate the two effects in $\Omega$ consider the  following, simple calculation.  We performe it in all generality in coordination $z$ on a bipartite lattice of alternating vertices $A-B$. We take for simplicity $\kappa=0$ (and thus integration over currents enforces $\psi=0$)  and $\vec H=0$.  

$\Omega[\phi]$ in Eq.~(\ref{Omega-n})  can be written as products  on $A$ vertices $v_a$, or
\begin{equation}
\Omega[\phi] =2^{N_l}\prod_{v_a}\omega[\phi_{v_a}]
\label{Omega-n2}
\end{equation}
with
\begin{equation}
\omega[\phi_{v_a}] =\prod_{v_b \in \partial v_a} \cos(\phi_{v_a}- \phi_{v_b}),
\label{Omega-n3}
\end{equation}
where $\partial v_{a}$ is the set of vertices $B$, connected to $v_{a}$. 
The next step is to perform the mean field approximation $\cos(\phi_{v_a}- \phi_{v_b})\simeq\cos(\phi_{v_a}-\overline \phi_{v_a})$, where $\overline \phi_{v_a}$ is the mean of the fields $\phi_{v_b}$ on vertices $v_{b}$ neighboring $v_a$, so that $\omega[\phi_{v}] \simeq \cos(\phi_{v}-\overline \phi_{v})^z$. That leads to
\begin{equation}
\int d\phi_{v}e^{-\ii \phi_{v} q_{v}} \omega[\phi_{v} ]=e^{\ii \overline \phi_{v}q_{v} } \tilde \omega(q_{v})
\end{equation}
where
\begin{equation}
 \tilde\omega (q)=2^{-z}\sum_{n=0}^{z} {{z} \choose {n}} \delta(q-q_n)
\end{equation}
 restricts  $q$ to the only possible charges $q_n=(2n-z)$ for $n=-z, -z+1, \dots, z-1,z$, each with proper multiplicity ${{z} \choose {n}}$. 
 
We have thus obtained a mean field approach that, unlike the high $T$ approach, separates in $\Omega[\phi]$ the effect of the entropic interaction, now expressed via the mean field $\bar \phi$, from the enforcement of discreteness of charges, which is necessary at very low $T$ where excitations are sparse. 

Because at low $T$ the correlation length is much larger than the lattice constant, it is physically intuitive to take $i\overline \phi_v=\langle i\phi_{v} \rangle=V^e_{v}$, the entropic field, which is real. Then, the $A \leftrightarrow B$ symmetry and the previous equations imply that  the probability of a charge distribution $q_v$ can be factored as
\begin{align}
\rho[q]=\prod_v \rho_v(q_v),
\end{align}
in terms of the probability of having a charge $q$ on a vertex $v$, given by
\begin{equation}
\rho_v(q)=\tilde \omega(q)  \exp\left(-\frac{ \epsilon}{T}q^2 + q V^e_{v}\right)/Z_v. 
\label{minkia}
\end{equation}

In Eq.~(\ref{minkia}) correlations are transmitted among vertices by the entropic field $V^e_{v}$. $Z_v$ normalizes $\rho_v(q)$ and depends on $v$ via $V^e_{v}$, which in turns depends on the {\em collective} charge distribution and is therefore non-local, or $V^e_{v}=V^e_{v}[q]$. Finally, while Eq.~(\ref{minkia}) was here deduced from a mean field approximation of the exact Eq.~(\ref{Z2-n}), it can however also be obtained directly {\em \`a la Landau}, so to speak, as we have shown previously~\cite{nisoli2014dumping}.

What we are missing now is an information on how the entropic field depends on the charge distribution. Assuming that at low $T$ charges are sparse,  that correlation lengths are large, and that the coarse graining of the geometry is isotropic (as it is the case for the square lattice), we can write the equation at lowest order in charges, fields and derivatives
\begin{equation}
\overline q =\chi_0 \nabla^2 V^e
\end{equation}
where $\overline q_v=\int dq \rho_v(q) q$. 
 From it and Eq.~(\ref{minkia}) we leave it as a simple exercise for the reader to show that linearization \`a la Debye-H\"uckel leads to the same correlations and screening as in the previous subsection, but with $\xi_{||}$ replaced by Eq.~(\ref{DH}) (see also ref~\cite{nisoli2020equilibrium}). 

\end{appendices}

\acknowledgements

We thank  Andrew King (D-Wave Systems) for useful discussions and Beatrice Nisoli for proofreading.
This work was carried out under the auspices of the U.S.
DoE through the Los Alamos National
Laboratory, operated by
Triad National Security, LLC
(Contract No. 892333218NCA000001). \\

{\bf Data Availability Statement}

Data sharing is not applicable to this article as no new data were created or analyzed in this study.

\bibliography{library2.bib}{}

\end{document}